\pdfoutput=1

\documentclass[11pt,nofootinbib]{article}
\usepackage{amsmath,amssymb,mathtools}
\usepackage{graphicx}
\usepackage[colorlinks=true,linkcolor=blue,citecolor=blue]{hyperref}
\usepackage{subfigure}
\usepackage{comment}
\usepackage{tensor}

\numberwithin{equation}{section}

\begin{document}

\providecommand{\abs}[1]{\lvert#1\rvert}
\providecommand{\bd}[1]{\boldsymbol{#1}}

\begin{titlepage}

\setcounter{page}{1} \baselineskip=15.5pt \thispagestyle{empty}

\begin{flushright}
SISSA 24/2016/FISI
\end{flushright}
\vfil


\bigskip
\begin{center}
 {\LARGE \textbf{Cosmological Aspects of Spontaneous Baryogenesis}}
\vskip 15pt
\end{center}

\vspace{0.5cm}
\begin{center}
{\Large 
Andrea De Simone}
and
{\Large 
Takeshi Kobayashi
}\end{center}

\vspace{0.3cm}

\begin{center}
\textit{SISSA, Via Bonomea 265,
 34136 Trieste, Italy}\\ 

\vskip 14pt
\textit{INFN, Sezione di Trieste, Via Bonomea 265,
 34136 Trieste, Italy}\\

\vskip 14pt
E-mail: 
\texttt{\href{mailto:andrea.desimone@sissa.it}{andrea.desimone@sissa.it}}, 
\texttt{\href{mailto:takeshi.kobayashi@sissa.it}{takeshi.kobayashi@sissa.it}}

\end{center} 



\vspace{1cm}

\noindent
We investigate cosmological aspects of spontaneous baryogenesis driven
 by a scalar field, and present general constraints that are independent
 of the particle physics model. The relevant constraints are obtained
 by studying the backreaction of the produced baryons on the
 scalar field, the cosmological expansion history after
 baryogenesis, and the baryon isocurvature perturbations.
 We show that cosmological considerations alone provide powerful
 constraints, especially for the minimal scenario with a quadratic scalar potential.
 Intriguingly, we find that for a given inflation scale,
 the other parameters including the reheat temperature, decoupling
 temperature of the baryon violating interactions, and the mass and
 decay constant of the scalar are restricted to lie within ranges of at
 most a few orders of magnitude. We also discuss possible extensions to
 the minimal setup, and propose two ideas for evading constraints on
 isocurvature perturbations: one is to suppress the baryon isocurvature
 with 
 nonquadratic scalar potentials, another is to compensate the
 baryon isocurvature with cold dark matter isocurvature by making the scalar survive
 until the present.

\vfil

\end{titlepage}

\newpage
\tableofcontents

\section{Introduction}
\label{sec:intro}

One of the great mysteries of our universe is the origin of the excess
of matter over antimatter.
The three basic ingredients for creating baryon number in the early
universe was laid out by Sakharov~\cite{Sakharov:1967dj}, and since then
various proposals for baryogenesis have been 
put forward. Among them, the idea of spontaneous
baryogenesis~\cite{Cohen:1987vi} is unique 
in the sense that it does not require Sakharov's third condition of a
departure from thermal equilibrium. Instead, it invokes spontaneous
breaking of the $CPT$ symmetry, which shifts the energy of baryons
relative to that of antibaryons, thus allowing baryon number production
even in equilibrium.

A general class of theories of spontaneous baryogenesis involves a
scalar field that is derivatively coupled to the baryon current
as~$(\partial_\mu \phi ) j_{B}^\mu$.
When the scalar field velocity~$\partial_0 \phi$ can be treated as a classical
background, the $CPT$ symmetry is spontaneously broken and thus a baryon
asymmetry can be produced.
For implementation of this mechanism in particle physics
models, see e.g. \cite{Cohen:1988kt,Dine:1990fj,Cohen:1991iu,Dolgov:1994zq,Dolgov:1996qq,Li:2001st,Chiba:2003vp,Carroll:2005dj,Kusenko:2014lra,Kusenko:2014uta,Daido:2015gqa}.
It should be noted that a theory of spontaneous baryogenesis does
not have to be specifically designed to yield a 
velocity~$\partial_0 \phi$ that does not vanish when averaged over
space. 
This is because cosmic
inflation~\cite{Starobinsky:1980te,Sato:1980yn,Guth:1980zm} can 
support spontaneous baryogenesis by providing a coherent motion of the
$\phi$-condensate that gives rise to a net baryon number in the observable
universe.
In this sense, the mechanism of spontaneous baryogenesis fits well into
inflationary cosmology.

However we note that a consistent embedding of
spontaneous baryogenesis into the early universe 
calls for a careful examination of the dynamics of the $\phi$~field.
This includes the dynamics during baryogenesis;
here one has to verify whether there is significant backreaction
from the produced baryons on~$\phi$. 
The fate of the $\phi$-condensate after baryogenesis is also important.
After creating the baryons the $\phi$ field oscillates
about its potential minimum and behaves as pressureless dust;
if $\phi$ dominates the universe before decaying, it would dilute the
baryon asymmetry, as well as impact the cosmological expansion history.
Furthermore, the dynamics of~$\phi$ before baryogenesis also has
observational consequences,
as during inflation the $\phi$~field obtains fluctuations which lead to
baryon isocurvature perturbations,
as was originally pointed out in~\cite{Turner:1988sq}.
There are now strong bounds on isocurvature perturbations from
measurements of the comic microwave background (CMB), which add further
conditions to be met by a successful spontaneous baryogenesis scenario.
These cosmological constraints on the $\phi$~field
have not necessarily been fully
taken into account in models studied in previous works,
and thus a systematic analysis of the cosmology with spontaneous
baryogenesis is required in order to verify the validity of the mechanism.

In this paper we consider the broad class of theories of spontaneous
baryogenesis driven by a scalar field derivatively coupled to
baryon currents.
By investigating the backreaction of the generated baryons on the scalar,
the expansion history of the universe after baryogenesis,
and the baryon isocurvature perturbations, we present general conditions
for spontaneous baryogenesis to create the baryon asymmetry in our
universe. 
In particular, for the minimal model with a quadratic potential for the
scalar field, we show that cosmological constraints alone restrict the
model parameters, including the inflation scale, to lie within a rather
narrow window.

We also suggest possible directions for extending the minimal setup,
to introduce new possibilities for spontaneous baryogenesis. 
For example, it has been known that the isocurvature constraint makes it
difficult for spontaneous baryogenesis to be compatible with high-scale
inflation. 
This issue is made sharper in this paper by combining
the constraints on isocurvature perturbations with various other
conditions. 
However, we also propose some ideas for ameliorating this issue with the
aid of a nonquadratic potential for the scalar. 
In particular, we point out that the baryon isocurvature perturbations can be
suppressed for linear potentials, or potentials with inflection points
such as cosine potentials.
We also discuss a rather exotic but topical possibility that the baryon
isocurvature perturbations are compensated by cold dark matter (CDM) isocurvature
perturbations, which can happen if the oscillating scalar is allowed to
survive until the present and constitute (a fraction of) the
CDM.\footnote{Some previous works also discussed ways to
suppress baryon isocurvature perturbations.
For example,
\cite{Kusenko:2014lra} considered stabilizing the scalar in a false
vacuum during inflation.
Issues with baryon isocurvature can also be evaded when spontaneous
baryogenesis is driven by domain walls~\cite{Daido:2015gqa}, or if the
baryon current is coupled instead to a derivative of the Ricci scalar,
i.e. $(\partial_\mu R) j_B^\mu $~\cite{Davoudiasl:2004gf}.\label{foot1}}

The paper is organized as follows: We start in Section~\ref{sec:SB}
by giving a brief review of spontaneous baryogenesis and setting our
notations. We then discuss the scalar dynamics  
during and after spontaneous baryogenesis in
Section~\ref{sec:during} and Section~\ref{sec:after}, respectively.
General discussions on the baryon isocurvature perturbations are
presented in Section~\ref{sec:baryon-iso}.
We then summarize the constraints for spontaneous baryogenesis in
Section~\ref{sec:sum-case}, and also illustrate the considerations
with a case study of a minimal model.
In Section~\ref{sec:nonquad} we discuss possible extensions to the
minimal scenario in order to alleviate constraints on baryon
isocurvature. Finally, we conclude in Section~\ref{sec:conc}.
We also provide comments and bounds on spontaneous baryogenesis induced
by the decay of the scalar in Appendix~\ref{app:A}.

\section{Brief Review of Spontaneous Baryogenesis}
\label{sec:SB}

Let us start by reviewing the basic mechanism of spontaneous
baryogenesis~\cite{Cohen:1987vi}.
This section also serves to set our notation.

\subsection{Basic Setup}
\label{subsec:setup}

As we mentioned in the introduction, in this paper we examine
spontaneous baryogenesis driven by a real scalar field~$\phi$ with a
derivative coupling to the baryon current,
\begin{equation}
 S = \int d^4 x \sqrt{-g} \left\{
  -\frac{1}{2} g^{\mu \nu} \partial_\mu \phi \partial_\nu \phi - V(\phi)
  - \sum_i c_i \frac{\partial_\mu \phi}{f}j_i^\mu
  +  \cdots
	 \right\}.
 \label{action}
\end{equation}
Here $j_i^\mu$ represents the current of a particle/antiparticle
pair~$i$,
whose baryon number is $B_i$ for the particle and $-B_i$ for the
antiparticle.
The time component of the current represents the difference
in the number density between the particle and antiparticle,
i.e., $j_i^0 = n_i - \bar{n}_i$.
For example, the current could be $j^\mu = \bar{q} \gamma^\mu q$ with
the quarks~$q$.
The sum $\sum_i$ runs over all particle species coupled to~$\phi$.
Moreover, $c_i$ is a dimensionless constant, $f$ is a mass scale,
and the dots represent terms that are independent of~$\phi$. 
Typically, $\phi$ would be a pseudo-Nambu-Goldstone boson (PNGB) of an
approximate symmetry corresponding to the baryon number,
and $f$ would be the associated symmetry breaking scale
(e.g. \cite{Cohen:1987vi,Dolgov:1994zq,Dolgov:1996qq}.)
In this paper, to keep the discussions general, we do not specify the
identity of~$\phi$ beyond what appears in the (effective) Lagrangian~(\ref{action}).
However with a slight abuse of language, we will refer to~$f$ as the
``decay constant.''

\begin{figure}[t!]
  \begin{center}
  \begin{center}
  \includegraphics[width=0.53\linewidth]{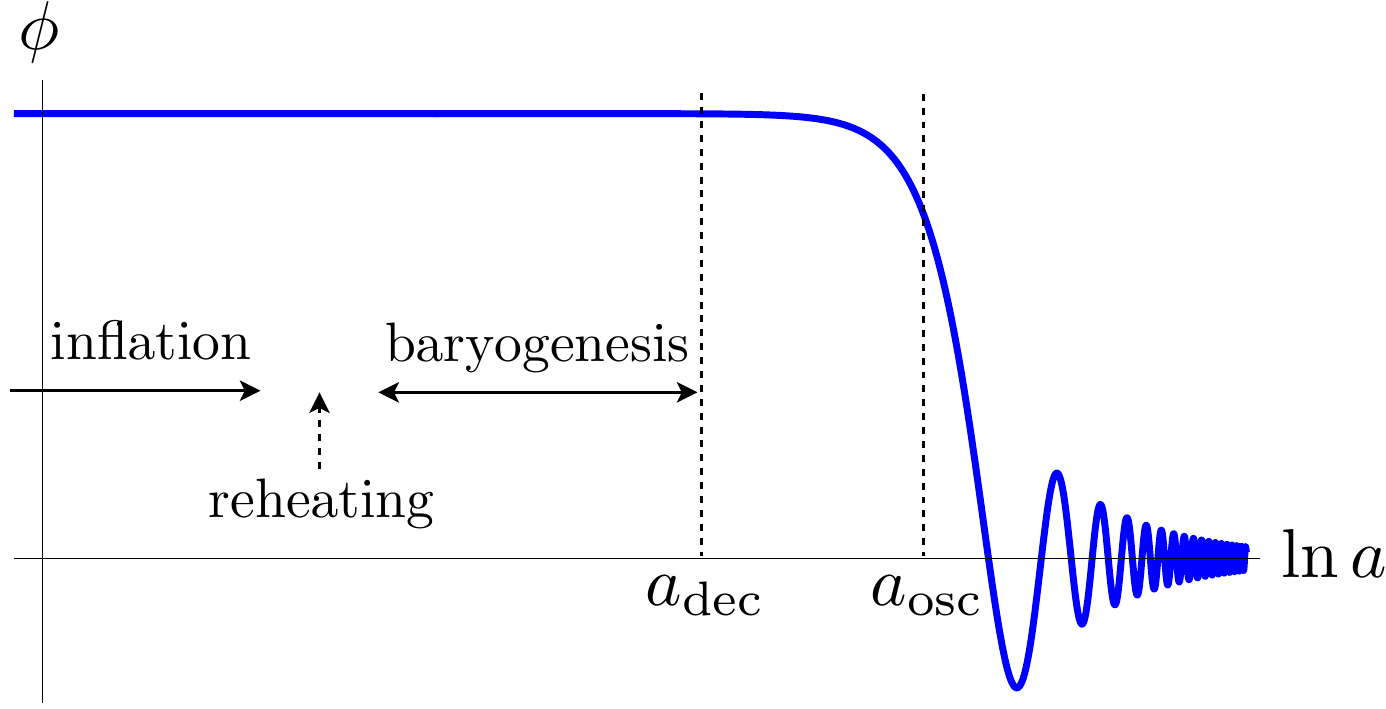}
  \end{center}
  \caption{Schematic of the scalar field dynamics in spontaneous
   baryogenesis (not to scale).}
  \label{fig:schematic}
  \end{center}
\end{figure}

With the action~(\ref{action}), spontaneous baryogenesis proceeds as
follows (each stage will be discussed in more detail in the subsequent
sections): 
\begin{enumerate}
 \item Cosmic inflation sets $\phi$ to be (almost) spatially homogeneous
       throughout the observable universe.
 \item After inflation, the universe eventually undergoes reheating and
       becomes dominated by radiation. Supposing some baryon number
       nonconserving processes to be in equilibrium, the baryon
       asymmetry is produced through the $(\partial_\mu \phi)
       j^\mu$~term. 
 \item The baryon number nonconserving processes eventually fall out of
       equilibrium, and from then on the baryon number freezes in.
       We use $T_{\mathrm{dec}}$ to represent the decoupling temperature
       for the baryon violating interactions,
       where the subscript ``dec'' will also be used for any quantity
       evaluated at decoupling. 
 \item The scalar~$\phi$ has been slowly rolling along its potential
       while the baryon asymmetry was being produced. After
       decoupling, as the Hubble 
       friction becomes weaker, $\phi$ begins to oscillate
       about its potential minimum. We denote values at the onset of the
       $\phi$ oscillation by the subscript~``osc.''
 \item The oscillating~$\phi$ eventually decays away
       through the $(\partial_\mu \phi) j^\mu $~term.
\end{enumerate}
See Figure~\ref{fig:schematic} for a schematic of the $\phi$-dynamics.

Note that in the case where $\phi$ is identified as the PNGB, the
symmetry breaking should happen prior to inflation, which 
indicates that $f$ should be larger than the inflationary Hubble
rate and the reheat temperature.
Otherwise, unless with a specifically designed potential~$V(\phi)$,
the scalar velocity~$\partial_0 \phi$ would be close to zero when
spatially averaged over the observable universe, resulting only in an
extremely tiny baryon asymmetry.

We should also remark that baryon number can also be produced
while the coherent oscillation of~$\phi$ decays,
as was originally pointed out in~\cite{Cohen:1987vi}. However this effect is
suppressed~\cite{Dolgov:1994zq,Dolgov:1996qq}.
Moreover, using the constraints discussed in the following sections, 
we show explicitly in Appendix~\ref{app:A} that the baryon asymmetry
from the decay of~$\phi$ is typically much smaller than that in
our universe.
Therefore, in this paper
we focus on the baryon number produced in equilibrium, while the
scalar~$\phi$ is slowly rolling. 

\subsection{Energy Shift}

The effect of the spontaneous $CPT$ breaking is clearly seen in the
energy density sourced by the $\phi$-related terms in the 
action. Here, let us express the three terms in~(\ref{action}) as
$S_\phi  = \int d^4x \sqrt{-g} L_\phi$ .

When the particle species~$i$ are all bosons
(such as
$j_i^\mu = i ( \varphi_i^* \partial^\mu \varphi_i - \varphi_i
\partial^\mu \varphi_i^* )$ with complex scalars~$\varphi_i$),
the energy-momentum tensor is obtained by varying
the action in terms of the metric,
\begin{equation}
 T_{\mu \nu} = g_{\mu \nu} L_{\phi} - 2 \frac{\partial
  L_{\phi}}{\partial g^{\mu \nu }} . 
\end{equation}
Then, considering a flat FRW background
\begin{equation}
 ds^2 = -dt^2 + a(t)^2 d \bd{x}^2,
\end{equation}
with a homogeneous~$\phi$, i.e.,
\begin{equation}
 \phi = \phi(t), 
\end{equation}
the energy density is obtained as
\begin{equation}
 \rho = -\tensor{T}{_0^0}  = 
  \frac{1}{2} \dot{\phi}^2 + V(\phi)
  - \sum_i c_i \frac{\dot{\phi}}{f} j_i^0,
  \label{energy-density}
\end{equation}
where we use an overdot to denote a derivative in terms of the
cosmological time~$t$.

In the case where the species~$i$ are fermions (such as
$j_i^\mu = \bar{\psi_i} \gamma^\mu \psi_i$), 
we vary the action in terms of a vierbein $e^{a \mu}$
(here the metric is constructed as $g_{\mu \nu} = \eta_{ab} e^a_{\mu}
e^b_\nu$, and $a$, $b$ are the indices in the local Minkowski space.)
Rewriting the fermion current as $j^\mu = e^{a \mu} j_a$, we can compute
the energy-momentum tensor as
\begin{equation}
 T_{\mu \nu} = g_{\mu \nu} L_\phi - e^a_\nu \frac{\partial
  L_{\phi}}{\partial e^{a \mu}}, 
\end{equation}
which, under a homogeneous background yields
\begin{equation}
  \rho = -\tensor{T}{_0^0}  = 
 - L_\phi + \dot{\phi}^2 - \sum_i c_i
  \frac{\dot{\phi}}{f} j_{i}^0. 
\end{equation}
Here, after imposing the equations of motion of the fermion fields,
the $\dot{\phi}j_i^0$ terms included in $L_{\phi}$ should be
cancelled by other fermion contributions to the energy density
(i.e. those arising from $(\cdots)$ in the action~(\ref{action}).)
Hence we obtain the same expression for the
energy density as we did for the boson currents in~(\ref{energy-density}). 

For both boson and fermion currents, the coupling to
$\partial_\mu \phi$ gives
a contribution to the energy density of 
$- c_i (\dot{\phi} / f) j_i^0 = - c_i (\dot{\phi} / f) (n_i -
\bar{n}_i)$.
This indicates that when the $\phi$ field can be considered as a classical background,
the spontaneous $CPT$ breaking assigns
for each particle/antiparticle pair~$i$ an extra energy of
\begin{equation}
 \Delta E_i = -c_i \frac{\dot{\phi}}{f}
  \label{Eshift}
\end{equation}
per particle, and $-\Delta E_i$ per antiparticle.
When in equilibrium, this can
alternatively be interpreted as particles obtaining a chemical
potential of
\begin{equation}
 \mu_i = c_i \frac{\dot{\phi} }{f},
  \label{chpot}
\end{equation}
while $-\mu_i$ for antiparticles.

\subsection{Baryon Asymmetry}

Supposing the particles and antiparticles to be in thermal equilibrium
after reheating, and 
also baryon number nonconserving processes to be occurring rapidly,
then the energy shift~(\ref{Eshift}) gives rise to a baryon asymmetry. 
Hereafter we assume all the species~$i$ to be relativistic fermions, and
ignore their masses.\footnote{In the case with boson species, if the induced
energy shift~$\abs{\Delta E}$ is larger than their masses, then naively,
the bosons are expected to be produced explosively,
until they back react on~$\phi$ and slow down the velocity.
It would be interesting
to understand explicitly what happens in such cases.}  
Further supposing there are no other symmetries that restrict their
thermal distributions,
then the difference in the number densities between the particles and
antiparticles is computed from the chemical potential~(\ref{chpot}) as
\begin{equation}
 \begin{split}
 j_i^0 = n_i - \bar{n}_i &= \frac{g_i}{(2\pi)^3}
  \int d^3 p
\left[ 
 \left\{\exp \left( \frac{p-\mu_i}{T}  \right)  + 1 \right\}^{-1}
  -   \left\{\exp \left( \frac{p+\mu_i}{T}  \right)  + 1 \right\}^{-1}
  \right] \\
    &= \frac{g_i}{6} \mu_i T^2
  \left\{  1 + \mathcal{O}\left(\frac{\mu_i}{T} \right)^2
  \right\}.
  \label{j0comp}
 \end{split}
\end{equation}
Here $g_i$ is the internal degrees of freedom of the
(anti)particle~$i$,
and in the second line we carried out an expansion in terms of the ratio
$\mu_i / T$, assuming it to be tiny, 
\begin{equation}
\left( \frac{\mu_i}{T}\right)^2 \ll 1.
\label{small_mu}
\end{equation}
Thus a baryon asymmetry has been produced, with number density
\begin{equation}
 n_B = \sum_i B_i (n_i - \bar{n}_i) =
    \sum_i \frac{B_i  c_i  g_i}{6} \frac{T^2\dot{\phi}}{f}  .
  \label{n_B}
\end{equation}

The baryon number freezes in when the baryon nonconserving
processes fall out of equilibrium.
Thus its ratio to the entropy density
\begin{equation}
 s = \frac{2 \pi^2}{45}g_{s* } T^3,
  \label{s-T}
\end{equation}
freezes at the value upon decoupling of the baryon violating
interactions as
\begin{equation}
  \left. \frac{n_B}{s}  \right|_{\mathrm{dec}} =
 \left.
\frac{15}{4 \pi^2}\frac{\sum_i B_i c_i g_i }{g_{s*}}\frac{\dot{\phi}}{T
f}
		       \right|_{\mathrm{dec}}.
 \label{nB-to-s-inT}
\end{equation}
If there is no further baryon nor entropy production afterwards,
and neglecting sphaleron processes which give order-unity corrections,  
then this ratio remains constant and so should match the present day
value $(n_B / s)_0 \approx 8.6 \times 10^{-11}$
measured by {\it Planck}~\cite{Ade:2015xua}.
(We use the subscript~``$0$'' to denote values today.)

\section{Dynamics of $\phi$ During Baryogenesis}
\label{sec:during}

We now move on to investigate the dynamics of the scalar field~$\phi$.
We start by discussing the dynamics during baryogenesis in this
section.

It should be stressed that,
for the baryon asymmetry to be spontaneously generated in thermal equilibrium,
the $\phi$~field should not starts its oscillation until the 
baryon violating processes decouple.
In terms of the Hubble rate $H = \dot{a}/a$, this condition is
written as
\begin{equation}
 H_{\mathrm{dec}} > H_{\mathrm{osc}}.
  \label{oscafterdec}  
\end{equation}
Otherwise, at the time of decoupling, which is
when the time scale of the baryon violating interaction~$\tau_{B}$
becomes comparable to the Hubble time,
the $\phi$ field would be oscillating with a time period much shorter
than~$\tau_{B}$, as $ \tau_B \sim H_{\mathrm{dec}}^{-1} \gg m_\phi^{-1}$. 
Then the chemical potential should be obtained as (\ref{chpot})
averaged over the oscillations, and would vanish.

Here we note that the decay of the oscillating~$\phi$ can also produce a
baryon number, which may even start from the early stages of the
oscillations. 
However such effects are likely to be tiny, as we discuss in Appendix~\ref{app:A}.

Therefore, we consider the scalar field to be slowly
varying along its potential while the baryons are being
produced.\footnote{Since the slow variation of the $\phi$-condensate is
required after reheating, $\phi$ could not have played the role of the
inflaton.}

\subsection{Slow-Varying Attractor}

The homogeneous equation of motion of~$\phi$ is obtained from the
action~(\ref{action}) as
\begin{equation}
 \ddot{\phi}  + 3 H \dot{\phi} +V'(\phi) = \sum_i 
 \frac{c_i}{f}
  \frac{ \partial_t (a^3 j_i^0)}{a^3},
  \label{EoM}
\end{equation}
where we have dropped the spatial components of the currents.
Let us first ignore the source term in the right hand side.
As the spontaneous baryogenesis takes place after reheating,
here we are interested in a radiation-dominated universe, where
\begin{equation}
 \frac{\dot{H}}{H^2} = -2.
  \label{HdotRD}
\end{equation}
Then while the field's effective mass is smaller than the Hubble rate as
\begin{equation}
 \left|  \frac{V''(\phi)}{5 H^2} \right| \ll 1,
  \label{slow-var}
\end{equation}
the equation of motion is approximated by
\begin{equation}
 5 H \dot{\phi} \simeq -V'(\phi).
  \label{slowRD}
\end{equation}
This is an attractor solution which is similar to the 
inflationary slow-roll approximation, except for that the numerical
coefficient in the left hand side is a $5$ instead of a~$3$. 
This is due to the time-dependence of the Hubble
friction in a radiation-dominated universe~(\ref{HdotRD}); 
in fact, $\phi$ accelerates on this attractor as
\begin{equation}
 \ddot{\phi} \simeq 2 H \dot{\phi},
  \label{ddotphi}
\end{equation}
as is clearly seen by comparing 
(\ref{slowRD}) with the original~(\ref{EoM}).
For detailed discussions on attractor solutions in an expanding
universe, see, e.g., \cite{Chiba:2009sj} or Appendix~A
of~\cite{Kawasaki:2011pd}. 

The relation~(\ref{ddotphi}) sets the rate of change of the 
chemical potential~(\ref{chpot}) as
$\mu _i / \dot{\mu}_i = (2 H)^{-1} $.
Hence we see that while the time scale for the baryon violating
interactions~$\tau_B$ is shorter than the Hubble time,
the chemical potential varies slow enough so that the particles and
antiparticles can follow their thermal distributions
of~(\ref{j0comp}).  

Using the slow-varying approximation~(\ref{slowRD}),
and rewriting the temperature of the radiation-dominated universe
in terms of the Hubble rate as
\begin{equation}
 \frac{\pi^2}{30}g_* T^4 = \rho_{\mathrm{r}} = 3 M_p^2 H^2 ,
  \label{TtoH}
\end{equation}
where $ \rho_{\mathrm{r}}$ is the radiation energy density,
then the baryon number density~(\ref{n_B}) is expressed as
\begin{equation}
 n_B = -\frac{\sum_i B_i c_i g_i }{\pi \, (10 \, g_*)^{1/2}} 
  \frac{M_p V'(\phi)}{f},
  \label{nB5.1}
\end{equation}
and the final baryon-to-entropy ratio~(\ref{nB-to-s-inT}) becomes
\begin{equation}
 \left. \frac{n_B}{s} \right|_{\mathrm{dec}}
  = - \left( \frac{9}{2560 \, \pi^6} \right)^{1/4}
   \sum_i B_i c_i g_i \, 
 \frac{g_*^{1/4}(T_{\mathrm{dec}})}{g_{s*}(T_{\mathrm{dec}})}
 \frac{V'(\phi_{\mathrm{dec}})}{f M_p^{1/2} H_{\mathrm{dec}}^{3/2}}.
 \label{nB-to-s}
\end{equation}
Let us also rewrite the assumption of a tiny chemical
potential~(\ref{small_mu}) as
\begin{equation}
 \left( \frac{\mu_i}{T} \right)^2
  =   \frac{\pi c_i^2 g_*^{1/2}}{75 \sqrt{10}}
 \frac{ \left( V'(\phi) \right)^2}{f^2 M_p H^3 }
 \ll 1,
 \label{tiny-mu}
\end{equation}
which is now a condition on the shape of the scalar potential compared to the Hubble
rate.

\subsection{Backreaction}

Now let us estimate the right hand side of the 
equation of motion~(\ref{EoM}) which we have been ignoring,
\begin{equation}
 \sum_i 
 \frac{c_i}{f}
  \frac{ \partial_t (a^3 j_i^0)}{a^3} = 
  \frac{3 \sqrt{10}}{2 \pi }
  \frac{\sum_i  c_i^2 g_i}{g_*^{1/2}}\frac{M_p H}{f^2} H \dot{\phi} .
\end{equation}
Here we computed the time-derivative of the particle
density~(\ref{j0comp}) using (\ref{HdotRD}), 
(\ref{ddotphi}), and (\ref{TtoH}).
(We ignored the derivative of~$g_*$.)
As each of the terms in the left hand side of~(\ref{EoM}) is~$ \sim H
\dot{\phi}$, we see that the source term is negligible in the equation
of motion when 
\begin{equation}
 \frac{3 \sqrt{10}}{2 \pi }
  \frac{\sum_i  c_i^2 g_i}{g_*^{1/2}}\frac{M_p H}{f^2} \ll 1
  \label{neg-BR}
\end{equation}
is satisfied.
This is roughly the same as requiring the
cosmic temperature $T \sim (M_p H)^{1/2}$ to be smaller
than the decay constant~$f$.
Otherwise, if $ T \gtrsim f$, the backreaction from
the particles in the thermal bath would be relevant, which is likely to
slow down the scalar velocity and thus suppresses
baryogenesis. Note also that in the case where $\phi$ is a PNGB
of a symmetry corresponding to the baryon number,
the symmetry would be recovered if the temperature were high enough to
violate~(\ref{neg-BR});
this would spoil spontaneous baryogenesis as was 
discussed in Section~\ref{subsec:setup}.

\section{Dynamics of $\phi$ After Baryogenesis}
\label{sec:after}

After decoupling, the scalar field eventually starts to
oscillate as the Hubble friction becomes weaker.
The time when the field actually starts to oscillate depends on the
detailed form of the 
scalar potential~$V(\phi)$; hence to keep the discussions general,
we proceed by representing the field value at the onset of the
oscillations by~$\phi_{\mathrm{osc}}$. 
However, we suppose that $V(\phi)$ is well-approximated by a quadratic
around its minimum, 
and that the oscillation of~$\phi$ quickly settles down to a harmonic
once the oscillation begins. 
Then the energy density of the oscillating~$\phi$ can be estimated as
\begin{equation}
 \rho_\phi = V(\phi_{\mathrm{osc}})
  \left( \frac{a_{\mathrm{osc}}}{a} \right)^3,
  \label{7.1}
\end{equation}
where we ignored the kinetic energy of~$\phi$ at the beginning of the
oscillation. 
Since the oscillating $\phi$ behaves as pressureless dust, its energy
density relative to that of radiation grows in time.
On the other hand, the derivative coupling~$(\partial_\mu \phi) j^\mu $
provides the oscillating~$\phi$ with a decay channel. 
So the important question here is, does the $\phi$-condensate dominate
the universe before decaying away?

In the case where $\phi$ dominates the universe, the universe will have
to be heated up again (which is most likely to be initiated
by the decay of~$\phi$), in order to connect to the standard Hot Big
Bang cosmology.
Hence the produced baryon asymmetry would be diluted
by the entropy production during the second reheating.
 
We also remark that, since $\phi$ has super-horizon field fluctuations
obtained during inflation, a $\phi$ that dominates or comes close to
dominating the universe would produce adiabatic density perturbations
{\it \` a la}
curvatons~\cite{Linde:1996gt,Enqvist:2001zp,Lyth:2001nq,Moroi:2001ct}. 
However, such adiabatic perturbations produced after baryogenesis should
only be a small fraction of the entire adiabatic perturbations in order to
keep the baryon isocurvature within observational
bounds;\footnote{Although, we also mention that the baryon isocurvature
due to the curvaton-like~$\phi$ can in principle be cancelled by the
isocurvature produced during spontaneous baryogenesis,
as both perturbations originate from the same
field fluctuations of~$\phi$. 
If such cancellation happens, and if sufficiently large baryon asymmetry still
remains after the second reheating, then $\phi$ could be 
responsible for the generation of the baryon number as well as the
entire adiabatic perturbations in our universe.}
we shall come back to this point in the next section.

\subsection{Hypothetical Relic Abundance}

In order to see whether $\phi$ ever dominates the universe, 
let us temporarily assume that $\phi$ survives until the present, and
compute its relic abundance. 
From the entropy conservation $s \propto a^{-3}$, the $\phi$ density
(\ref{7.1}) can be rewritten as
\begin{equation}
 \rho_\phi = V(\phi_{\mathrm{osc}}) \frac{s}{s_{\mathrm{osc}}}.
  \label{rhophi-in-s}
\end{equation}
Then using the relation between the entropy and the radiation density
(cf.~(\ref{s-T}) and (\ref{TtoH})):
\begin{equation}
 s = \frac{2 \pi^2 g_{s*}}{45} \left( \frac{30 \,
\rho_\mathrm{r}}{\pi^2 g_*}      \right)^{3/4} 
\label{s-rho}
\end{equation}
to evaluate $s_{\mathrm{osc}}$, 
and considering the universe at the onset of the oscillation to still be
dominated by radiation, i.e., 
\begin{equation}
 V(\phi_{\mathrm{osc}})  \ll  \rho_{\mathrm{r\, osc}} \simeq 3 M_p^2
  H_{\mathrm{osc}}^2, 
  \label{rhoradosc}
\end{equation}
one obtains the hypothetical $\phi$ abundance today as
\begin{equation}
 \begin{split}
  \Omega_\phi h^2
  \equiv \frac{\rho_{\phi 0} \,   h^2}{3 M_p^2 H_0^2}
   &=   \frac{45}{2 \pi^2 g_{s*}(T_{\mathrm{osc}})} \left( \frac{\pi^2
  g_*(T_{\mathrm{osc}})}{30 \cdot 3 M_p^2 H_{\mathrm{osc}}^2} 
  \right)^{3/4  } 
    \frac{V(\phi_{\mathrm{osc}}) s_0 h^2}{3 M_p^2 H_0^2} \\
  & \approx 2.9 \times 10^{26} \, 
    \frac{g_*(T_{\mathrm{osc}})^{3/4}}{g_{s*}(T_{\mathrm{osc}})}
  \frac{V(\phi_{\mathrm{osc}}) }{ M_p^{5/2} H_{\mathrm{osc}}^{3/2} }.
  \label{Omegaphi}
  \end{split}
\end{equation}
If this $\Omega_\phi$ is larger than the
measured matter abundance~$\Omega_{\mathrm{m}}$,
then $\phi$ would dominate the universe before the standard
matter-radiation equality, unless it decayed at earlier times. 

Now let us further suppose $\Omega_\phi \gg \Omega_{\mathrm{m}}$,
and evaluate when $\phi$ would dominate the universe. 
Here, during the times between the onset of the oscillation and the
$\phi$-domination, we only need to consider $\phi$ and radiation as the
major components of the universe. 
Let us represent values at the hypothetical $\phi$-radiation
equality by the subscript~``dom,'' i.e.,
\begin{equation}
 \rho_{\phi\,  \mathrm{dom}} = \rho_{\mathrm{r}\, \mathrm{dom}} =
  \frac{3 M_p^2 H_{\mathrm{dom}}^2}{2}.
\end{equation}
Then combining this with (\ref{rhophi-in-s}), (\ref{s-rho}), and
(\ref{rhoradosc}), one finds 
\begin{equation}
 H_{\mathrm{dom}} = \frac{\sqrt{2}}{9}
  \left( \frac{g_* (T_{\mathrm{osc}})}{g_* (T_{\mathrm{dom}})} \right)^{3/2}
  \left( \frac{g_{s*} (T_{\mathrm{dom}})}{g_{s*} (T_{\mathrm{osc}})} \right)^2
  \frac{V^2(\phi_{\mathrm{osc}})}{M_p^4 H_{\mathrm{osc}}^{3}}.
  \label{Hdom}
\end{equation}

\subsection{Fate of $\phi$} 

We now restore $\phi$'s decay channels provided by the
$(\partial_\mu \phi) j^\mu / f$ coupling.
Let us parameterize the decay rate by
\begin{equation}
 \Gamma_{\phi}  = \beta \frac{m_{\phi}^3}{f^2},
  \label{Gamma_phi}
\end{equation}
where $m_\phi$ is the mass of $\phi$ around its potential minimum,
i.e. $  m_\phi^2 = V''(\phi_{\mathrm{min}}) $, and
$\beta$ is a dimensionless constant.
The explicit value of~$\beta$ depends on the particle physics model, but
is much smaller than unity in many cases. 
The derivative term $(\partial_\mu\phi) j^\mu$ in the Lagrangian
typically includes baryon couplings of the $\phi$ field, as well as
anomalous couplings to $W\tilde{W}$, $Z\tilde{Z}$.
The former becomes irrelevant when the baryon violating interactions go
out 
of equilibrium below $T_{\rm dec}$, so the latter would provide the most
important decay modes of $\phi\to WW, ZZ$, with
$\beta\sim 10^{-6}$.
Further decay channels could also exist if there are
other sources of baryon violation contributing to~$\partial_\mu
j^\mu$, or additional couplings in the Lagrangian other than the
derivative term.  
To keep our discussions general, we collectively describe all the decay
channels of~$\phi$ by the expression in~(\ref{Gamma_phi}), and we
proceed without specifying the value of~$\beta$.

Cosmological constraints on the decay rate depend on whether the would-be
abundance~$\Omega_\phi$~(\ref{Omegaphi}) is larger or smaller than the
CDM abundance~$\Omega_{\mathrm{CDM}}$.

\subsubsection*{Case with $\Omega_\phi > \Omega_\mathrm{CDM}$ :}

If the would-be $\phi$ abundance exceeds the CDM abundance,
then $\phi$ obviously needs to decay prior to the standard
matter-radiation equality in order not to spoil the Big Bang expansion
history.
Moreover, the $\phi$~density at the time of Big Bang Nucleosynthesis is 
strictly restricted, so it is preferable for $\phi$ to decay before then.

However even if $\phi$ decays long before nucleosynthesis, if it had
dominated the universe before decaying, then the already produced baryon
asymmetry would have been greatly diluted; see discussions
below~(\ref{7.1}). 
In order to avoid $\phi$ from dominating the universe in the first
place, the condition
\begin{equation}
 \Gamma_\phi > H_{\mathrm{dom}}
  \label{non-dom}
\end{equation}
is required, where $H_{\mathrm{dom}}$ was obtained
in~(\ref{Hdom}).\footnote{Strictly speaking, $H_{\mathrm{dom}}$ in
(\ref{Hdom}) was obtained 
assuming $\Omega_{\phi} \gg \Omega_\mathrm{m}$;
so in the case of
$\Omega_{\mathrm{m}} \gtrsim \Omega_\phi > \Omega_{\mathrm{CDM}}$,
the condition~(\ref{non-dom}) should be corrected by
a factor of order unity.}

\subsubsection*{Case with $\Omega_\phi \leq \Omega_\mathrm{CDM}$ :}

In this case the $\phi$ density is guaranteed to be subdominant at least
until the matter-radiation equality.
If further $\Omega_\phi$ is close to~$\Omega_{\mathrm{CDM}}$, then
$\phi$ should either decay prior to the standard matter-radiation
equality so as not to drastically modify the Big Bang evolution,
or alternatively $\phi$ may survive until today and constitute (a fraction of) CDM.
The latter possibility is discussed in Section~\ref{subsec:CIP}.

\vspace{\baselineskip}

Before closing this section, we should also comment on the thermal
production of~$\phi$.
We have analyzed the energy density of the $\phi$-condensate,
however, a thermal distribution of the $\phi$-particles can also be
produced through the $(\partial_\mu \phi) j^\mu$ term 
while the baryon violating processes are occurring rapidly,
depending on the nature of the processes.
The energy density of the relativistic $\phi$-particles in equilibrium would
be much larger than that of the $\phi$-condensate. 
After decoupling, the $\phi$-particles would 
fall out of equilibrium and redshift initially as radiation, and then
as pressureless dust after the temperature of the universe drops below the particle
mass. Hence the $\phi$-particles, if they start from a thermal
distribution, can dominate the universe long before the
$\phi$-condensate would, and more strongly suppress the baryon asymmetry.
Therefore the conditions in this section which were obtained by studying
the $\phi$-condensate, such as~(\ref{non-dom}), 
should be regarded as conservative bounds.

\section{Baryon Isocurvature Perturbations}
\label{sec:baryon-iso}

As was discussed at the beginning of Section~\ref{sec:during}, the
$\phi$ field starts its oscillation after the decoupling.
This indicates that during inflation, the effective mass of~$\phi$ is
much lighter than the Hubble rate, and therefore $\phi$ obtains
super-horizon field fluctuations. Such fluctuations source isocurvature
perturbations in the baryon density, 
as is clear from the expression~(\ref{nB5.1}).
Let us compute the baryon isocurvature perturbations for general
scalar potentials~$V(\phi)$ in this section.

The gauge-invariant isocurvature perturbation between the baryons and
photons
(see e.g. \cite{Lyth:2009zz} for a review)
is defined as
\begin{equation}
 S_{B \gamma } \equiv  \frac{\delta n_B}{n_B} - \frac{3}{4} \frac{\delta
  \rho_\gamma }{\rho_\gamma },
  \label{Bisodef}
\end{equation}
which is clearly conserved while the baryon number and photon
energy densities redshift locally as $n_B \propto \rho_{\gamma}^{3/4}
\propto a^{-3}$.
Hence let us consider $S_{B \gamma}$ 
at temperatures a bit below $T = 1\, \mathrm{MeV}$ when the universe
is dominated by the photons and the decoupled neutrinos; 
we refer to this radiation-dominated epoch as the ``late RD.''
Supposing there are no isocurvature perturbations between the
photons and neutrinos, then the uniform-photon density slicings in the
late RD epoch coincide with uniform-total density slicings.
Thus the baryon isocurvature perturbation~(\ref{Bisodef}) is rewritten as
\begin{equation}
S_{B \gamma} = \left. \frac{\delta n_B}{n_B} \right|_{\rho
 = \mathrm{const.}, \, \mathrm{late} \,  \mathrm{RD}}
 \label{SBg5.4}
\end{equation}
where the right hand side is evaluated on
a uniform-photon/total density slice.

In order to compute $S_{B \gamma}$ as a function of~$\phi$, we would
like to know how $\delta n_B / n_B$ during the late RD epoch relates to
that at the time of decoupling.
Here we consider the decoupling temperature~$T_{\mathrm{dec}}$ to be
a constant value set by the microphysical parameters
of the baryon violating interactions. Then, since the universe at
decoupling is 
also radiation-dominated, the decoupling surface where $ T =
T_{\mathrm{dec}}$ can be viewed as a uniform-density slice as well.

To see how $\delta n_B / n_B$
on super-horizon scales
evolves between two uniform-density slices, 
we consider the baryon number to be locally conserved since decoupling;
then the baryon number density at a comoving spatial coordinate~$\bd{x}$
satisfies 
\begin{equation}
 n_B (t_2, \bd{x}) =  n_B (t_1, \bd{x}) \left( \frac{a(t_1,
 \bd{x})}{a(t_2, \bd{x})}\right)^3,
\end{equation}
where $t_1$ and $t_2$ describe some arbitrary uniform-density slices
that are after decoupling. The scale factors on these slices are
written as 
\begin{equation}
 a(t_i, \bd{x}) = \bar{a}(t_i) e^{\zeta (t_i, \bd{x})},
\end{equation}
where the bar denotes the unperturbed value, 
$\zeta$ is the curvature perturbation on 
uniform-density slicings, and $i= 1, 2$.
Then one finds that, at the linear order in the
fluctuations around the unperturbed background,
\begin{equation}
 \frac{\delta n_B (t_2, \bd{x})}{\bar{n}_B (t_2)} =
  \frac{\delta n_B (t_1, \bd{x})}{\bar{n}_B (t_1)}
  + 3 \left(
 \zeta (t_1, \bd{x}) -  \zeta (t_2, \bd{x})
\right).
\end{equation}
Thus, taking one of the uniform-density slices to be during the late RD
epoch and the other one at decoupling, (\ref{SBg5.4}) becomes
(dropping the bars again for simplicity)
\begin{equation}
S_{B \gamma} = \left. \frac{\delta n_B}{n_B} \right|_{T
 = T_\mathrm{dec}} +
 3 \left(
    \zeta_{\mathrm{dec}} - \zeta_{\mathrm{late} \,  \mathrm{RD}}
	   \right).
\label{S-zeta}
\end{equation}
Here we have obtained the familiar result that any change in the curvature
perturbations after baryon number production gives rise to baryon
isocurvature perturbations. (This is the reason why $\phi$ should
not be a curvaton responsible for the entire adiabatic perturbations,
as was discussed below~(\ref{7.1}).)
In the following, we assume that $\zeta$ has approached its final value prior
to decoupling and drop the second term in~(\ref{S-zeta}).

The baryon number fluctuations arise from the field fluctuations of~$\phi$.
Hence from (\ref{nB5.1}), the baryon isocurvature is written as
\begin{equation}
 S_{B \gamma } =
    \frac{V''(\phi_{\mathrm{dec}})}{V'(\phi_{\mathrm{dec}})} \delta
    \phi_{\mathrm{dec}},
    \label{6.8}
\end{equation}
where $\delta \phi_{\mathrm{dec}}$ denotes the field fluctuation 
on the $T= T_{\mathrm{dec}}$ surface.
Considering~$\phi$ to have followed an attractor solution
of the sort discussed in (\ref{slowRD}) since during inflation,
then $\phi$'s field value (including fluctuations) at decoupling can be
viewed as a function of that on some initial flat slice during
inflation,
at around or after the modes of interest exited the
horizon.\footnote{The adiabatic perturbation~$\zeta$ may further source
field fluctuations on the decoupling surface, however we ignore
this effect assuming it to be tiny.}
Thus we rewrite (\ref{6.8}) as
\begin{equation}
  S_{B \gamma } =
    \frac{V''(\phi_{\mathrm{dec}})}{V'(\phi_{\mathrm{dec}})} 
    \frac{\partial \phi_{\mathrm{dec}} }{\partial \phi_{\mathrm{ini}}}
    \delta \phi_{\mathrm{ini}},
    \label{SBgreal}
\end{equation}
where $\phi_{\mathrm{ini}}$ denotes the unperturbed field value and
$\delta \phi_{\mathrm{ini}}$ the fluctuation
on an initial flat surface.

Let us now go to Fourier space, and take the initial flat surface as
when a pivot scale~$k_*$
exits the horizon during inflation. Denoting this time
by an asterisk, i.e. $k_* = a_* H_*$, then the
Fourier component of the isocurvature is
\begin{equation}
 ({S}_{B \gamma})_{ \bd{k}_* } =
    \frac{V''(\phi_{\mathrm{dec}})}{V'(\phi_{\mathrm{dec}})} 
    \frac{\partial \phi_{\mathrm{dec}} }{\partial \phi_{*}}
    (\delta \phi_{*})_{ \bd{k}_*} .
\label{SBgfourier}
\end{equation}
From the power spectrum of the field fluctuations upon horizon
exit,\footnote{The power spectrum $\mathcal{P}(k)$ is defined as
\begin{equation}
 \langle \delta \phi_* (\bd{x}) \delta \phi_* (\bd{y}) \rangle
  = \int \frac{d^3 k}{4 \pi k^3}e^{i \bd{k}\cdot (\bd{x} - \bd{y})}
  \mathcal{P}_{\delta \phi_*} (k).
\end{equation}
}
\begin{equation}
 \mathcal{P}_{\delta \phi_*} (k_*) = \left( \frac{H_*}{2 \pi} \right)^2,
  \label{phiH2pi}
\end{equation}
we arrive at our final expression for the baryon isocurvature
perturbation spectrum,
\begin{equation}
 \mathcal{P}_{B \gamma} (k_*) =
  \left( 
  \frac{V''(\phi_{\mathrm{dec}})}{V'(\phi_{\mathrm{dec}})} 
  \frac{\partial \phi_{\mathrm{dec}} }{\partial \phi_{*}}
  \frac{H_*}{2 \pi} \right)^2.
  \label{P_Bgamma}
\end{equation}
This expression which is a function of~$\phi$ is the main result of this
section. We will explicitly compute the right hand side in later
sections when we discuss specific examples.
The isocurvature spectrum is nearly scale-invariant,
as $\phi$'s effective mass is much lighter than the Hubble rate
during inflation. 
Furthermore, supposing $\phi$ not to contribute to 
the adiabatic perturbations, then the baryon isocurvature is
uncorrelated with the adiabatic perturbation.

Isocurvature perturbations are well constrained by measurements of the
CMB anisotropies.
However, since CMB does not distinguish between baryon and CDM
isocurvature modes at linear order, the baryon isocurvature is
constrained as an effective CDM isocurvature, 
\begin{equation}
 \mathcal{P}_{\mathrm{CDM}  \gamma}^{\mathrm{eff}} (k)=
    \left( \frac{\Omega_B}{\Omega_{\mathrm{CDM}}} \right)^2
    \mathcal{P}_{B \gamma} (k)
    \approx 0.034  \times   \mathcal{P}_{B \gamma} (k) .
    \label{effCDMiso}
\end{equation}
The {\it Planck} limit on a scale-invariant and uncorrelated
isocurvature perturbation reads~\cite{Ade:2015lrj}
\begin{equation}
 \mathcal{P}^{\mathrm{eff}}_{\mathrm{CDM} \gamma} (k_*)
  \lesssim   0.040 \times 
  \mathcal{P}_{\zeta} (k_*)
  \quad
  (95\%\, \mathrm{C.L.}, \, \, \mathrm{TT,TE,EE+lowP}  )
 \label{iso-con}  
\end{equation}
on the pivot scale $k_*/ a_0 = 0.05 \,
\mathrm{Mpc}^{-1}$,
where the adiabatic power is 
$ \mathcal{P}_{\zeta} (k_*) \approx 2.2 \times 10^{-9} $.

\section{Summary of Constraints and a Case Study}
\label{sec:sum-case}

We now put together the constraints discussed in the previous sections.

\subsection{Generic Conditions for Spontaneous Baryogenesis}
\label{subsec:gencon}

The generic conditions for spontaneous baryogenesis to operate are as
follows: 

During spontaneous baryogenesis, the $\phi$ field is required
not to start its oscillation 
until after decoupling~(\ref{oscafterdec}),
and in particular its effective mass should be lighter than the
Hubble rate~(\ref{slow-var}). Furthermore, the backreaction from
the produced particles, or the thermal friction, has to be
suppressed~(\ref{neg-BR}) for sufficient baryogenesis.
Then, assuming a tiny chemical potential~(\ref{tiny-mu}),
the final baryon-to-photon ratio is computed as~(\ref{nB-to-s}), which
should give the present day value of 
$(n_B / s)_0 \approx 8.6 \times 10^{-11}$. 

After baryogenesis, the $\phi$ density is supposed to be subdominant
at least until the onset of the oscillations~(\ref{rhoradosc}).
In the case where the hypothetical
$\phi$~abundance~(\ref{Omegaphi}) exceeds the CDM abundance,
if one would like to avoid the baryon asymmetry from being diluted and
thus desires to prevent $\phi$ from dominating the universe,
then $\phi$ is required to decay before domination~(\ref{non-dom}), where
$H_{\mathrm{dom}}$ is given in~(\ref{Hdom}). 
This condition becomes stronger
if a thermal distribution of $\phi$-particles is produced during baryogenesis, 
as discussed at the end of Section~\ref{sec:after}.

Another restriction comes from the baryon isocurvature
perturbations~(\ref{P_Bgamma}), which is constrained by {\it Planck}
as~(\ref{iso-con}).  

We should also remark that there are further constraints if $\phi$ is a
PNGB.
As was discussed towards the end of Section~\ref{subsec:setup}, 
the symmetry breaking needs to happen prior to inflation, thus
\begin{equation}
 f > H_*.
  \label{symbreak}
\end{equation}
For the PNGB case, the backreaction condition~(\ref{neg-BR}) can also be
understood as the requirement that thermal fluctuations do not recover
the symmetry.
The field range of a PNGB~$\phi$ is also restricted by the decay
constant~$f$, typically as 
\begin{equation}
 \abs{\phi_*  - \phi_{\mathrm{min}}  } \lesssim f,
  \label{field-range}
\end{equation}
where $\phi_{\mathrm{min}}$ is the minimum of the scalar potential. 
The explicit field range bound, e.g. whether it is $f$ or $\pi f$,
depends on the individual models. 
We also note that the field bound can apply not only for PNGBs but for
general cases,
as from an effective field theory point of view, the
violation of~(\ref{field-range}) would, at least naively, indicate a
breakdown of the perturbative description.\footnote{However it may be
possible to extend the field range while controlling the corrections to
the effective action by, for instance, invoking
monodromy~\cite{Silverstein:2008sg,McAllister:2008hb}.}

\subsection{Case Study: Quadratic Potential}
\label{subsec:CS-quad}

Having laid out the general conditions, in this subsection we study
how they actually constrain the minimal model with a quadratic
potential, 
\begin{equation}
 V(\phi) = \frac{1}{2} m_\phi^2 \phi^2.
    \label{quadV}
\end{equation}
If, for example, $\phi$ is a PNGB of an approximate U(1) symmetry, 
then quadratic potentials are realized in the vicinity of 
one of the minima of the periodic potential.

For quadratic potentials, the Hubble rate at the onset of oscillations
can simply be estimated as when the slow-varying
condition~(\ref{slow-var}) breaks down, i.e.,
\begin{equation}
 H_{\mathrm{osc}} = \frac{m_\phi }{\sqrt{5}}.
  \label{Hosc-quad}
\end{equation}
Since $\phi$ varies only slowly while on the
attractor~(\ref{slowRD}), we make the approximation of
\begin{equation}
 \phi_* \simeq \phi_{\mathrm{dec}} \simeq \phi_{\mathrm{osc}}. 
  \label{samephi}
\end{equation}
This in particular gives $ \partial \phi_{\mathrm{dec}} / \partial
\phi_* \simeq 1$ in the expression for the isocurvature
spectrum~(\ref{P_Bgamma}).\footnote{An alternative
definition of the onset of oscillations was given 
in~\cite{Kawasaki:2011pd} as
when the field variation during one Hubble time becomes
comparable to the distance to the potential minimum, i.e.
\begin{equation}
 \left| \frac{\dot{\phi}}{H (\phi - \phi_{\mathrm{min}})}
			    \right|_{\mathrm{osc}} = 1.
  \label{osc-def}
\end{equation}
For a quadratic potential, this definition combined with the
slow-varying approximation~(\ref{slowRD}) gives the 
same result as~(\ref{Hosc-quad}). 
However for nonquadratic potentials, $H_{\mathrm{osc}}$ given by 
(\ref{osc-def}) generally does not coincide with that estimated as when 
(\ref{slow-var}) is violated.
The use of $H_{\mathrm{osc}}$ and $\phi_{\mathrm{osc}}$ defined
by~(\ref{osc-def}) is suitable for accurate calculations of density
perturbations with nonquadratic potentials, as was demonstrated in
e.g.~\cite{Kawasaki:2011pd,Kobayashi:2013nva}.} 
In the following analyses we further impose the field bound, 
discussed around~(\ref{field-range}), as
\begin{equation}
 \abs{\phi_*} < f.
\label{quad9}  
\end{equation}
Here we have fixed the field bound to~$f$, however, changing the bound
by an order-unity factor (say, to~$\pi f$) gives only minor corrections
to our discussions below. 

As for the particle content of the theory, 
we consider the number of species~$i$ to be of order unity, with
parameters of order unity as well, i.e., $B_i \sim c_i \sim g_i \sim 1$.
The relativistic degrees of freedom $g_*$, $g_{s*}$
in the early universe are considered to be of $\sim 100$.
We bear in mind these numerical values in the following analyses,
although for completeness, we explicitly display the dimensionless
constants as well 
as the ratio $\abs{\phi_*} / f$ in the results.
For this purpose, let us introduce the normalized relativistic degrees
of freedom
\begin{equation}
 \tilde{g}_{*} \equiv \frac{g_*}{100}, \qquad
 \tilde{g}_{s*} \equiv \frac{g_{s*}}{100}.
\end{equation}

\subsubsection{Constraints}

The requirement of slow-variation of~$\phi$ until decoupling, 
cf.~(\ref{slow-var}), reads
\begin{equation}
 \frac{m_\phi^2}{5 H_{\mathrm{dec}}^2} \ll 1,
  \label{quad2}
\end{equation}
from which $H_{\mathrm{dec}} > H_{\mathrm{osc}}$~(\ref{oscafterdec}) is
automatically satisfied. 
On the other hand, the condition for negligible
backreaction~(\ref{neg-BR}), evaluated at decoupling, gives
\begin{equation}
 0.2 \times \frac{\sum_i c_i^2 g_i}{\tilde{g}_{*\mathrm{dec}}^{1/2}}
  \frac{M_p H_{\mathrm{dec}}}{f^2} \ll 1.
  \label{quad3}
\end{equation}
Recall that the sum~$\sum_i$ runs over all particles species
coupled to~$\phi$. 
Normalizing the baryon-to-photon ratio~(\ref{nB-to-s}) to the
present day value $(n_B / s)_0 \approx 8.6 \times 10^{-11}$ 
(ignoring sphaleron processes) yields
\begin{equation}
   \sum_i B_i c_i g_i
 \frac{\tilde{g}_{* \mathrm{dec}}^{1/4} }{\tilde{g}_{s*\mathrm{dec}}}
 \frac{m_\phi^2 \, \phi_*}{f M_p^{1/2} H_{\mathrm{dec}}^{3/2}} \approx
 -6  \times 10^{-8}.
    \label{quad4}
\end{equation}
Under this constraint, it can be checked that the assumption of a 
tiny chemical potential~(\ref{tiny-mu}) is satisfied. 
One can also obtain a lower bound on the decoupling temperature by 
solving (\ref{quad4}) for~$m_{\phi}$ and substituting it
into (\ref{quad2}), which yields
\begin{equation}
 H_{\mathrm{dec}} \gg
  400\, \mathrm{GeV} \times
  \frac{1}{(\sum_i B_i c_i g_i)^2}
 \frac{\tilde{g}_{s*\mathrm{dec}}^{2} }{\tilde{g}_{*\mathrm{dec}}^{1/2}}
 \left( \frac{f}{\phi_*} \right)^2 .
  \label{quad2-4}
\end{equation}
Note here that the $(f/ \phi_*)^2$ factor in the right hand side is
at least of order unity due to the field bound~(\ref{quad9}). 
The existence of the lower bound on the decoupling scale can be
understood from the fact that
a somewhat large mass~$m_\phi$ is required in order to provide 
the sufficient $\phi$~velocity for creating the baryon asymmetry, while
$H_{\mathrm{dec}}$ should be even larger than~$m_\phi$ to prevent $\phi$
from oscillating. 

The constraints can be combined to further
give a lower bound on the hypothetical
$\phi$~abundance~(\ref{Omegaphi}),
\begin{align}
 \Omega_\phi h^2  & \approx 2 \times 10^{26} \,
 \frac{\tilde{g}_{*\mathrm{osc}}^{3/4} 
  }{\tilde{g}_{s*\mathrm{osc}} }
 \frac{m_\phi^{1/2} \phi_*^2}{M_p^{5/2}}
  \label{Omegaphi-quad}
 \\
  & \gg 70 \times \frac{\tilde{g}_{*\mathrm{osc}}^{3/4}
  \,  \tilde{g}_{s*\mathrm{dec}}^3
 }{\tilde{g}_{s*\mathrm{osc}} 
  \, \tilde{g}_{*\mathrm{dec}}^{5/4 }
 }
  \frac{\sum_i c_i^2 g_i}{|\sum_j B_j c_j g_j |^3}
 \frac{f}{\abs{\phi_*}},
 \label{quadOmegaphi}
\end{align}
where upon moving to the second line,
we first substituted (\ref{quad4}) for $\phi_*$,
then used the inequalities (\ref{quad2}) and (\ref{quad3})
respectively for $m_\phi$ and $f$,
and finally used the lower bound~(\ref{quad2-4}) for
$H_{\mathrm{dec}}$. 
We clearly see that if $\phi$ did not decay, it would readily
overclose the universe.\footnote{However, we should also remark that it
is in principle possible to suppress $\Omega_\phi $
below~$\Omega_{\mathrm{CDM}}$, by 
having a large number of particle species~$i$ of more than $\sim 100$. We will 
discuss this possibility in Section~\ref{subsubsec:EX-CIP}.}
Thus in order to avoid $\phi$ from dominating the universe and diluting
the baryon asymmetry, we require $\phi$ to decay before domination, 
cf.~(\ref{non-dom}), which yields a condition
\begin{equation}
 \beta
 \left(\frac{\tilde{g}_{*\mathrm{dom}} }{\tilde{g}_{*\mathrm{osc}}}
	      \right)^{3/2}
 \left( \frac{\tilde{g}_{s*\mathrm{osc}} }{\tilde{g}_{s*\mathrm{dom}} }
 \right)^2 
\frac{m_\phi^2 M_p^4}{f^2 \phi_*^4} \gtrsim 0.4.
  \label{quada1}
\end{equation}
Here, to be conservative, we have ignored the possibility of the thermal
production of $\phi$ during baryogenesis. 

On the other hand, the constraint on baryon isocurvature perturbation,
cf. (\ref{P_Bgamma}) and (\ref{iso-con}), yields
\begin{equation}
 \left( \frac{H_*}{\phi_*} \right)^2
  \lesssim 1 \times 10^{-7}.
    \label{quad7}
\end{equation}
This combined with the field bound~(\ref{quad9})
requires $f$ to be much larger than the inflationary Hubble rate,
and thus in the case where $\phi$ is a PNGB, the symmetry breaking is guaranteed to
have happened before inflation, cf.~(\ref{symbreak}). 
On can also check by combining
(\ref{quad3}), (\ref{quad4}), (\ref{quada1}), and (\ref{quad7}) that,
unless $\beta$ is much greater than unity, 
then $\abs{\phi_*} \ll M_p$;
therefore the condition~(\ref{rhoradosc}) for the $\phi$ density to be 
subdominant at the onset of oscillations is satisfied. 
The isocurvature constraint further sets an upper bound on the inflation scale; 
substituting (\ref{quad4}) for~$m_\phi$ into (\ref{quada1}), and further
combining with (\ref{quad7}) yields
\begin{equation}
 H_* \lesssim 2 \times 10^{12} \, \mathrm{GeV} \, 
 \left(
\frac{\beta}{|\sum_i B_i
  c_i g_i |}
  \frac{
  \tilde{g}_{s*\mathrm{dec}} \, 
  \tilde{g}_{*\mathrm{dom}}^{3/2} \, 
  \tilde{g}_{s*\mathrm{osc}}^{2} 
  }{
  \tilde{g}_{*\mathrm{dec}}^{1/4} \, 
  \tilde{g}_{*\mathrm{osc}}^{3/2} \, 
  \tilde{g}_{s*\mathrm{dom}}^{2}
  } 
			  \right)^{2/9}
   \left( \frac{\abs{\phi_*}}{f}  \right)^{2/9}
   \left( \frac{H_{\mathrm{dec}}}{H_*} \right)^{1/3},
   \label{quadii}
\end{equation}
where it should be noted that the factor $H_{\mathrm{dec}} / H_*$ in the right hand
side is smaller than unity.
Alternatively, one can combine (\ref{quad3}), (\ref{quad4}),
(\ref{quada1}), and (\ref{quad7}) to obtain a $\phi_*$-independent
bound, 
\begin{equation}
 H_* \lesssim 3 \times 10^{12} \, \mathrm{GeV} \, 
 \left(
\frac{\beta}{|\sum_i B_i c_i g_i | (\sum_j c_j^2 g_j)^{1/2} }
  \frac{
  \tilde{g}_{s*\mathrm{dec}} \, 
  \tilde{g}_{*\mathrm{dom}}^{3/2} \, 
  \tilde{g}_{s*\mathrm{osc}}^{2} 
  }{
  \tilde{g}_{*\mathrm{osc}}^{3/2} \, 
  \tilde{g}_{s*\mathrm{dom}}^{2}
  } 
			  \right)^{1/4}
 \left( \frac{H_{\mathrm{dec}}}{H_*} \right)^{1/4}.
 \label{quadiii}
\end{equation}
Which of the two bounds (\ref{quadii}) and (\ref{quadiii}) is stronger
depends on the explicit choice of the model parameters.

\vspace{\baselineskip}

Thus we have discussed all the conditions laid out in
Section~\ref{subsec:gencon}.
To summarize our findings for the quadratic case,
it turns out that there are six independent
conditions under which the others are automatically satisfied;
these are the constraints from the field bound~(\ref{quad9}),
slow-variation of~$\phi$ until decoupling~(\ref{quad2}),
negligible backreaction~(\ref{quad3}),
normalization from the baryon-to-photon ratio today~(\ref{quad4}), 
requirement for $\phi$ to decay before dominating the universe~(\ref{quada1}),
and the limit on baryon isocurvature perturbation~(\ref{quad7}).
Combining the six conditions also yields bounds
on the decoupling~(\ref{quad2-4}) and inflation scales (\ref{quadii}),
(\ref{quadiii}); 
when ignoring the coefficients and supposing $\beta \lesssim 1$, 
the bounds read roughly as 
\begin{equation}
 10^2 \, \mathrm{GeV} \ll H_{\mathrm{dec}} < H_* \lesssim 10^{12}\,
  \mathrm{GeV}.
  \label{2-to-12}
\end{equation}
One can further check that, 
once the values of $H_{\mathrm{dec}}$ and $H_*$ are
chosen within the bounds, the scalar mass~$m_\phi$ and the
decay constant~$f$ are constrained to lie within a rather narrow window.
Let us now show this explicitly.

\subsubsection{Parameter Space}
\label{subsub:Pspace}

\begin{figure}[t!]
\centering
\subfigure[$H_* = 10^{5}\, \mathrm{GeV}$]{%
  \includegraphics[width=.45\linewidth]{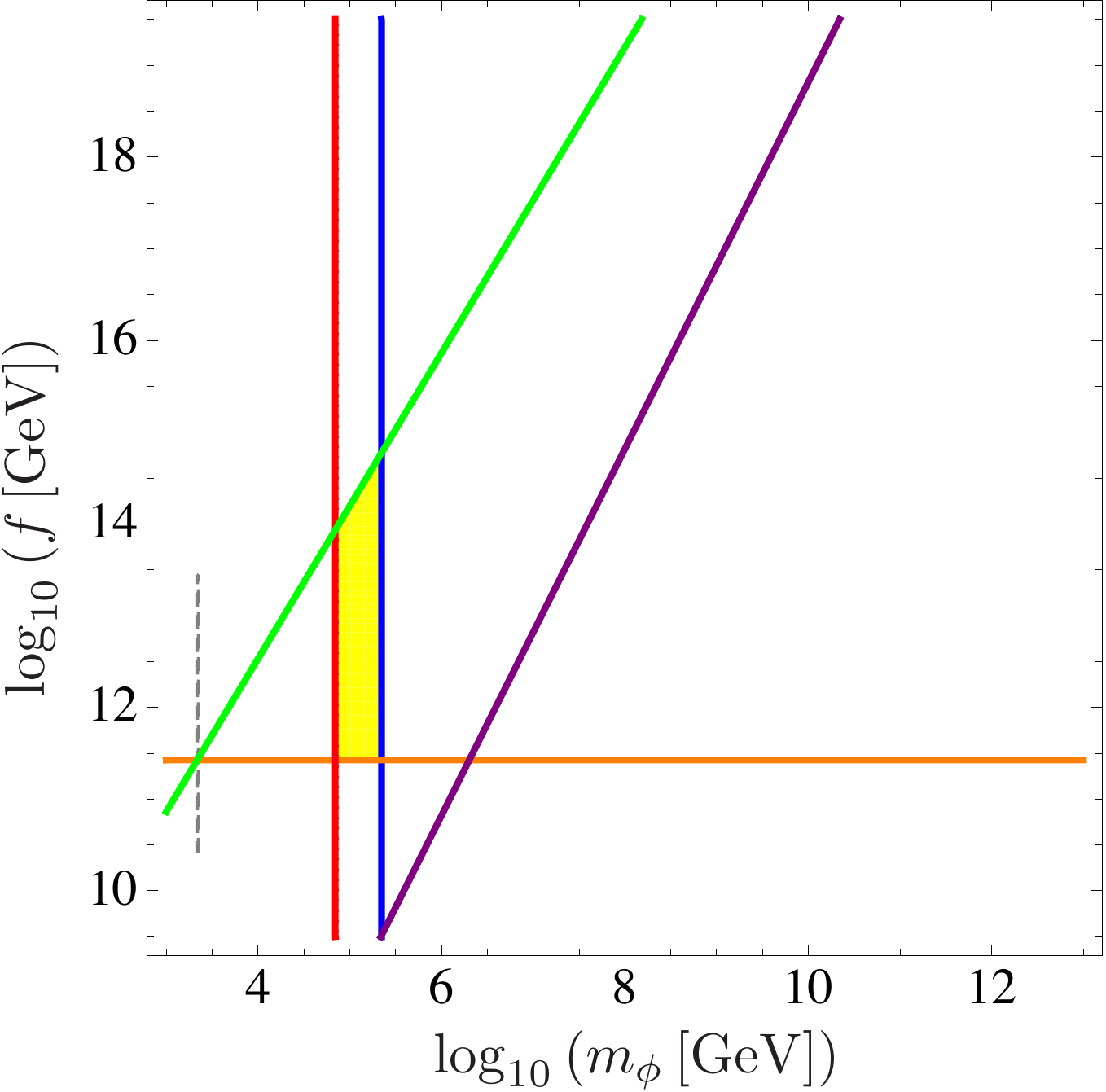}
  \label{fig:quad5}}
\quad
\subfigure[$H_* = 10^{7}\, \mathrm{GeV}$]{%
  \includegraphics[width=.45\linewidth]{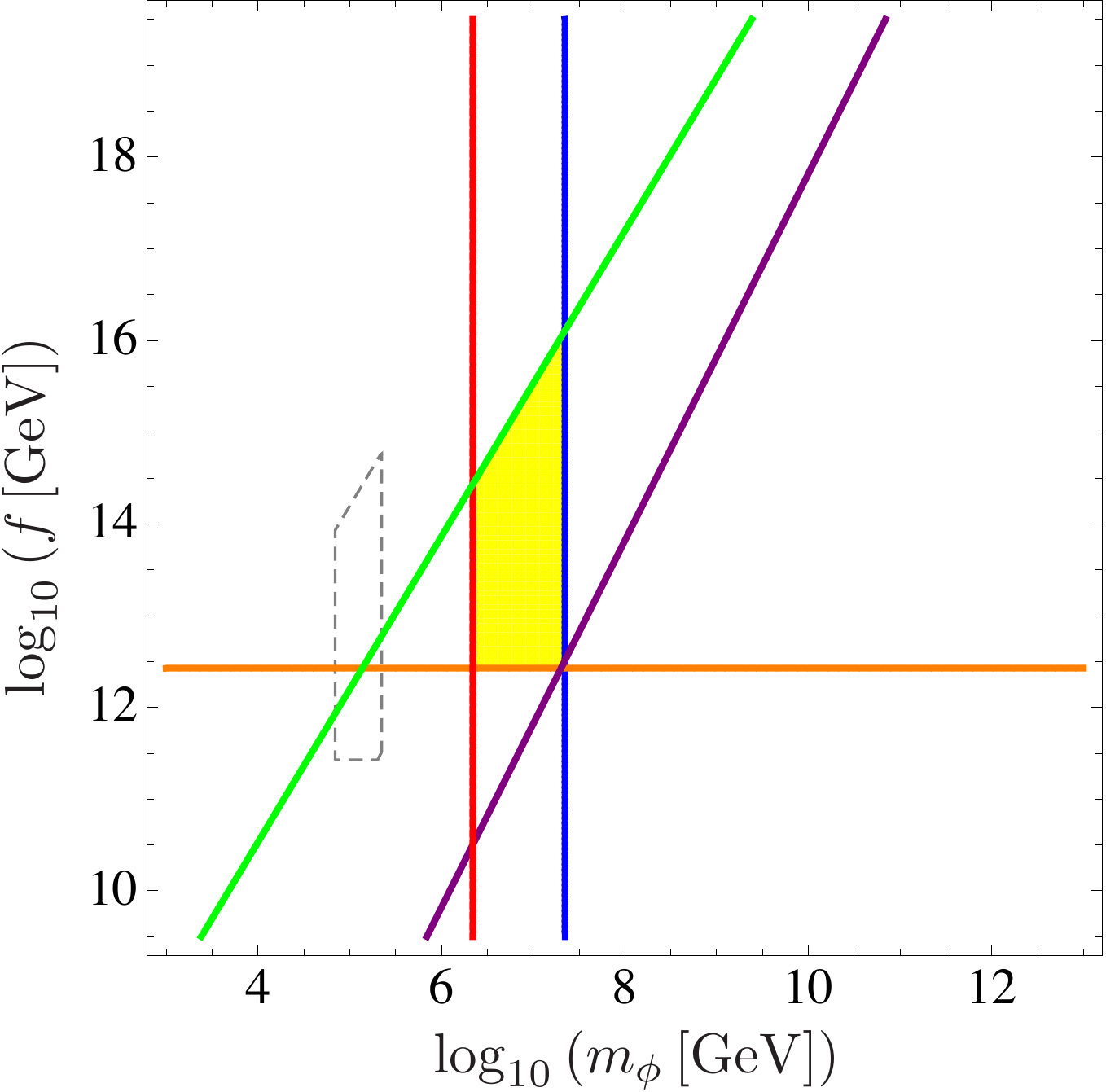}
  \label{fig:quad7}}
\subfigure[$H_* = 10^{9}\, \mathrm{GeV}$]{%
  \includegraphics[width=.45\linewidth]{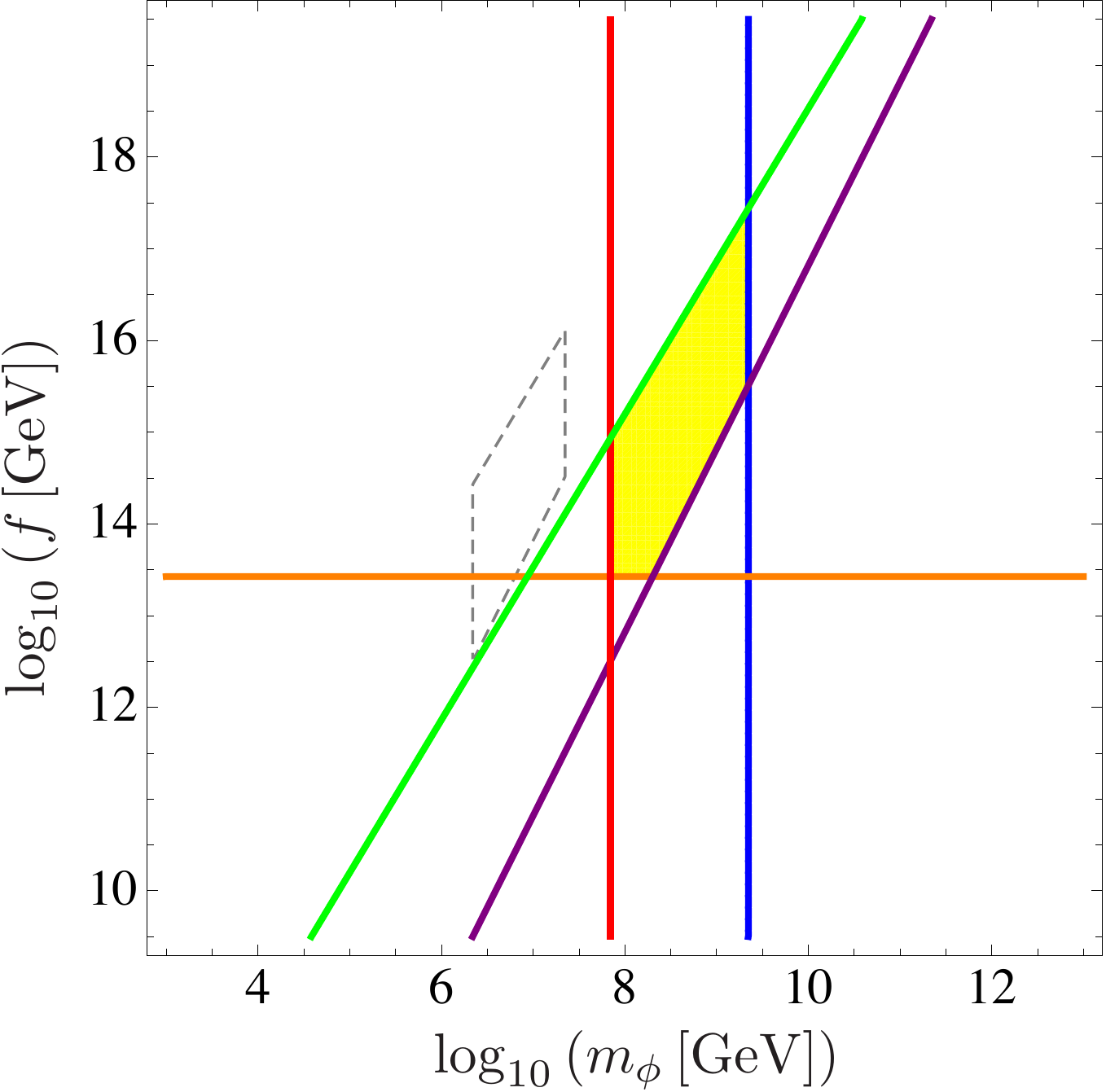}
  \label{fig:quad9}}
\quad
\subfigure[$H_* = 10^{11}\, \mathrm{GeV}$]{%
  \includegraphics[width=.45\linewidth]{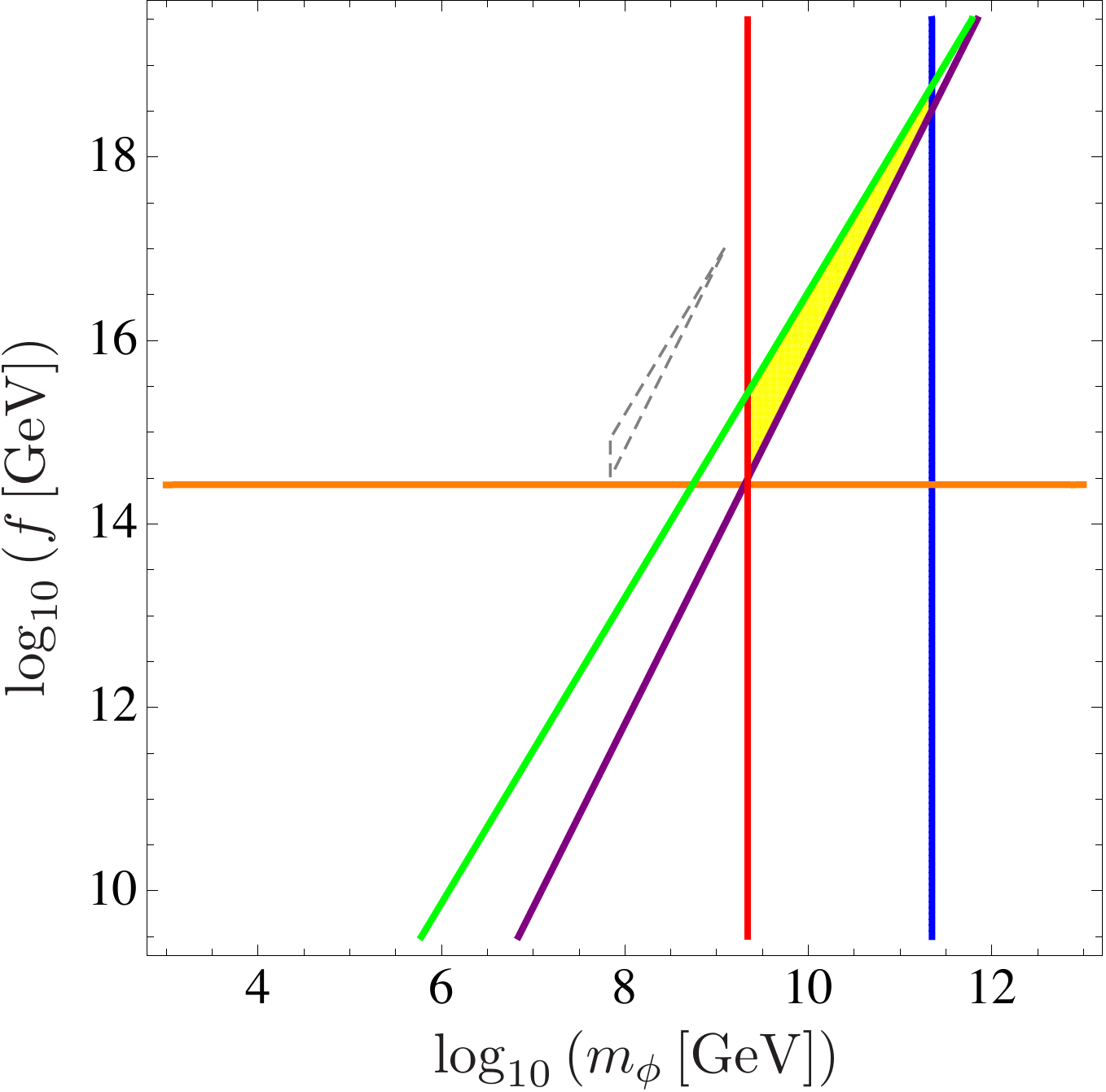}
  \label{fig:quad11}}
 \caption{Mass~$m_\phi$ and decay constant~$f$ of the scalar driving
 spontaneous baryogenesis with a quadratic potential. For each value of
 the inflation scale~$H_*$, the window for the parameters allowing
 sufficient baryogenesis is shown as the yellow region for the
 conservative case of $H_{\mathrm{dec}} = H_*$. The solid lines setting
 the boundaries represent the various constraints; 
 red: field bound~(\ref{quad9}), blue: slow-variation
 (non-oscillation) of scalar at decoupling~(\ref{quad2}), orange:
 insignificant backreaction from produced baryons~(\ref{quad3}),
 green: scalar decay before dominating the universe~(\ref{quada1}),
 purple: limit on baryon isocurvature perturbation~(\ref{quad7}). 
 For comparison, windows for $H_{\mathrm{dec}} = 10^{-2}  H_*$ are also
 shown as the regions bounded by gray dashed lines.}  
\label{fig:quadwindow}
\end{figure}

We present the window for the parameters in Figure~\ref{fig:quadwindow}.
Here, we have supposed there is one
species~$i$ coupled to~$\phi$, and  specified its parameters as
\begin{equation}
 B_i = \frac{1}{3}, \thickspace  c_i = 1, \thickspace   g_i = 2.
  \label{Bicigi}
\end{equation}
We also fixed the relativistic degrees of freedom at the times of decoupling, onset
of $\phi$ oscillation, and the hypothetical $\phi$ domination to the
total number in the Standard Model,
\begin{equation}
 g_* = g_{s*} = 106.75.
  \label{rdf106.75}
\end{equation}
As for the $\beta$~parameter for the decay rate~(\ref{Gamma_phi}),
note that a smaller~$\beta$ gives a smaller parameter window as it would delay the
time of decay and make it easier for $\phi$ to dominate the universe, 
cf.~(\ref{quada1}). For presenting conservative bounds, we fixed it to
\begin{equation}
 \beta = 1.
  \label{betaunity}
\end{equation}
Then there are five dimensionful parameters remaining, namely,
($H_{*}$, $H_{\mathrm{dec}}$, $m_\phi$, $f$, $\phi_*$). Among them, 
we fixed $\phi_*$ from the $n_B/s$~normalization~(\ref{quad4}),
and displayed the allowed window in the $m_\phi$-$f$ plane in
Figure~\ref{fig:quadwindow}.
We have chosen some values for the inflation scale from
within its bound, which now read
$ 0.9 \times 10^3\, \mathrm{GeV} \ll H_{\mathrm{dec}} < H_* \lesssim 2
\times 10^{12}\, \mathrm{GeV}$.
Here we remark that the allowed window becomes smaller
when there is a larger hierarchy between $H_{\mathrm{dec}}$ and $H_*$;
thus conservative bounds are obtained by assuming the two scales to be the same.

In the figures, the yellow regions denote the allowed window for the
conservative case of $H_{\mathrm{dec}} = H_*$.
The various constraints are represented by the solid lines
setting the boundaries;
red: (\ref{quad9}), blue: (\ref{quad2}), orange: (\ref{quad3}), 
green: (\ref{quada1}), purple: (\ref{quad7}).
(Recall that (\ref{quad4}) is used for fixing~$\phi_*$.)
It is firstly seen that, for each choice of the inflation scale,
$m_\phi$ and $f$ are constrained to lie within ranges of at most a few
orders of magnitude. 

The lower and upper bounds for~$m_\phi$ are mostly set by
the field bound (red) and the requirement
of slow-variation (blue). These two bounds approach each other
as $H_*$ is lowered, until they vanish the window; hence also
setting the lower bound on the inflation scale.
Note also from (\ref{quad4}) that $\abs{\phi_*} / f \propto m_\phi^{-2}$,
thus as one moves away from the red boundary towards larger~$m_\phi$, the
value of~$\abs{\phi_*}/ f$ becomes much smaller than unity.
There one would need to fine tune $\phi$'s initial position, unless there is
some dynamical mechanism that sets exactly the right value for~$\phi_*$. 

The decay constant~$f$, at low inflation scales,
is bounded from below by the backreaction condition (orange), and from
above by the requirement that $\phi$ does not dominate the universe (green).
As one goes to higher inflation scales,
the isocurvature constraint (purple) becomes relevant and eventually
eliminates the window.
When $\beta$ is smaller than unity, the green lines shift downwards
in the figures, and thus further shrink the windows,
although the dependence on~$\beta$ is not so strong.
After fixing~$\phi_*$ with (\ref{quad4}), the upper bound on~$f$ 
from~(\ref{quada1}) scales as $\propto \beta^{1/6}$;
thus e.g. when $\beta = 10^{-6}$, the green lines in the figures are
lowered by $\Delta (\log_{10} f) = -1$. 
We should
also remark that even if $\phi$ dominates, a sufficient
baryon asymmetry may remain if $\phi$ decays not so long after
dominating the universe. 
(Although one should also consider effects on the adiabatic
perturbations in this case, see discussions below~(\ref{7.1}).)
Thus the allowed window can actually extend beyond the green line to a certain
degree, however we do not expect this to drastically expand the parameter space.
(Note that a larger~$f$, while delaying the $\phi$ decay,
cf.~(\ref{Gamma_phi}), also suppresses the baryon number production,
cf.~(\ref{nB5.1}).)

We also plot windows for the case with $H_{\mathrm{dec}} = 10^{-2} H_*$
for comparison, as the regions bounded by gray dashed lines.
(In Figure~\ref{fig:quad5}, the dashed line close to the left edge
denotes a very thin band for~$m_\phi$.)
It can be clearly seen that the 
windows shrink when $H_{\mathrm{dec}}$ is lowered compared to~$H_{*}$. 
This indicates that, even though $H_*$ itself may vary over $10$~orders
of magnitude, cf.~(\ref{2-to-12}),
the energy scales of inflation, reheating, and decoupling all have to lie 
within a rather narrow range of, at the very most, a
few orders of magnitude. 
(Here, note also that $\rho^{1/4} \propto H^{1/2}$.)
Therefore an efficient (p)reheating is a prerequisite for a successful
spontaneous baryogenesis.

\section{Evading Isocurvature Constraints with Nonquadratic Potentials}
\label{sec:nonquad}

In the previous section we laid out the general conditions required for
spontaneous baryogenesis, then studied the minimal scenario with a
quadratic potential.
There we saw that spontaneous baryogenesis is strictly constrained by the CMB
limits on baryon isocurvature perturbations, especially with high inflation scales. 
In this section, we remark that the isocurvature constraints can be
alleviated by nonquadratic scalar potentials,
for instance with a cosine.
Such potentials actually do arise when $\phi$ is a PNGB of an
approximate U(1) symmetry; as then $V(\phi)$ is a periodic
potential, so one can imagine $\phi$ at the time of symmetry breaking to
roll down to a region away from the minima, where the potential cannot be
approximated by a quadratic. 

We discuss two possible solutions for evading the
isocurvature constraints; one is to suppress the baryon isocurvature,
and the other is to compensate the baryon and CDM isocurvature
perturbations.

\subsection{Suppressing Baryon Isocurvature}
\label{subsec:SBI}

From the expression~(\ref{P_Bgamma}), one notices that the baryon
isocurvature perturbation vanishes at linear order if
$V''(\phi_{\mathrm{dec}})$ is zero.
This is because of $\dot{\phi} \propto V'(\phi)$,
thus a vanishing second derivative of the potential makes the
$\phi$~velocity insensitive to the field value of~$\phi$.
In such cases, the large-scale field fluctuations do not lead to
inhomogeneities in the $\phi$~velocity at decoupling and thus no baryon
isocurvature modes are induced.
The simplest way to realize this is to consider a linear potential.
Alternatively, one can invoke potentials with inflection points.

In the following, with a PNGB~$\phi$ in mind, let us consider a cosine
potential of the form
\begin{equation}
 V(\phi) = m_\phi^2 f^2 \left[ 1 - \cos\left( \frac{\phi}{f} \right)
			   \right],
 \label{cosV}
\end{equation}
which asymptotes to the quadratic potential~(\ref{quadV}) in the
vicinity of $\phi = 0$.
As the cosine potential has the periodicity~$2 \pi f$ set by the
decay constant, we can focus on the field range 
$\abs{\phi} \leq \pi f$ without loss of generality.

The baryon isocurvature perturbation from the cosine potential 
can be analytically computed using~(\ref{SBgfourier}).
Considering $\phi$ to be effectively frozen until decoupling and thus using
the approximation of $\phi_{\mathrm{dec}} \simeq \phi_*$, we get
\begin{equation}
 ({S}_{B \gamma})_{ \bd{k}_* } =
      \tan^{-1} \left(\frac{\phi_*}{f} \right)
    \frac{(\delta \phi_*)_{\bd{k}_*}}{f}.
\label{SBgamma-cos}
\end{equation}
Let us split the Fourier mode
\begin{equation}
 ({S}_{B \gamma})_{ \bd{k} } =
 \sqrt{\frac{2 \pi^2}{k^3}} \, 
 ( \tilde{S}_{B \gamma})_{k} \, a_{\bd{k}}
 \label{Ssplit}
\end{equation}
into the amplitude $( \tilde{S}_{B \gamma})_{k}$ and
a stochastic variable~$a_{\bd{k}}$ that satisfies
$\langle a_{\bd{k}} a^*_{\bd{p}} \rangle = (2 \pi)^3 \delta (\bd{k} -
\bd{p}) $,
and likewise for the field fluctuation~$(\delta \phi_*)_{\bd{k}_*}$.
Note here that the square of the amplitude corresponds to the power spectrum,
\begin{equation}
 \mathcal{P}_{B \gamma} (k) = \left| ( \tilde{S}_{B \gamma} )_k \right|^2.
 \label{eqamppw}
\end{equation}
Thus (\ref{phiH2pi}) gives, up to an unimportant phase,
\begin{equation}
 (\widetilde{\delta \phi}_{*})_{k_*} = \frac{H_*}{2 \pi},
  \label{tildedeltaphih2pi}
\end{equation}
which is combined with (\ref{SBgamma-cos}) to yield
\begin{equation}
 ( \tilde{S}_{B \gamma})_{k_*} =
      \tan^{-1} \left(\frac{\phi_{*}}{f} \right)
      \frac{H_*}{2 \pi f}.
 \label{eq7.5}      
\end{equation}
One clearly sees that the baryon isocurvature is suppressed if
$\abs{\phi_{*}}/f$ is close to $\pi/2$.

We have also carried out numerical computations to check this behavior:
We numerically solved the equation of motion of~$\phi$~(\ref{EoM})
(ignoring the source term) in an FRW background universe, from the
inflationary epoch to the radiation-dominated epoch when decoupling happens.
By varying the initial position of~$\phi$ when the pivot scale exits the
horizon by $H_* / 2 \pi$,
we computed the resulting variation in the $\phi$~velocity at
decoupling, then evaluated the baryon isocurvature using
$S_{B \gamma} = (\delta \dot{\phi} / \dot{\phi})_{T =
T_{\mathrm{dec}}}$
(cf. (\ref{n_B}), (\ref{S-zeta}).)
Note that the baryon isocurvature computed in this way
corresponds to the amplitude of the Fourier mode~$ (\tilde{S}_{B \gamma})_{k_*} $,
defined in (\ref{Ssplit}) and analytically computed
as~(\ref{eq7.5}). 
Squaring the amplitude yields the 
the power spectrum at the pivot scale~$\mathcal{P}_{B \gamma} (k_*)$.

\begin{figure}[t!]
 \begin{minipage}{.48\linewidth}
  \begin{center}
 \includegraphics[width=\linewidth]{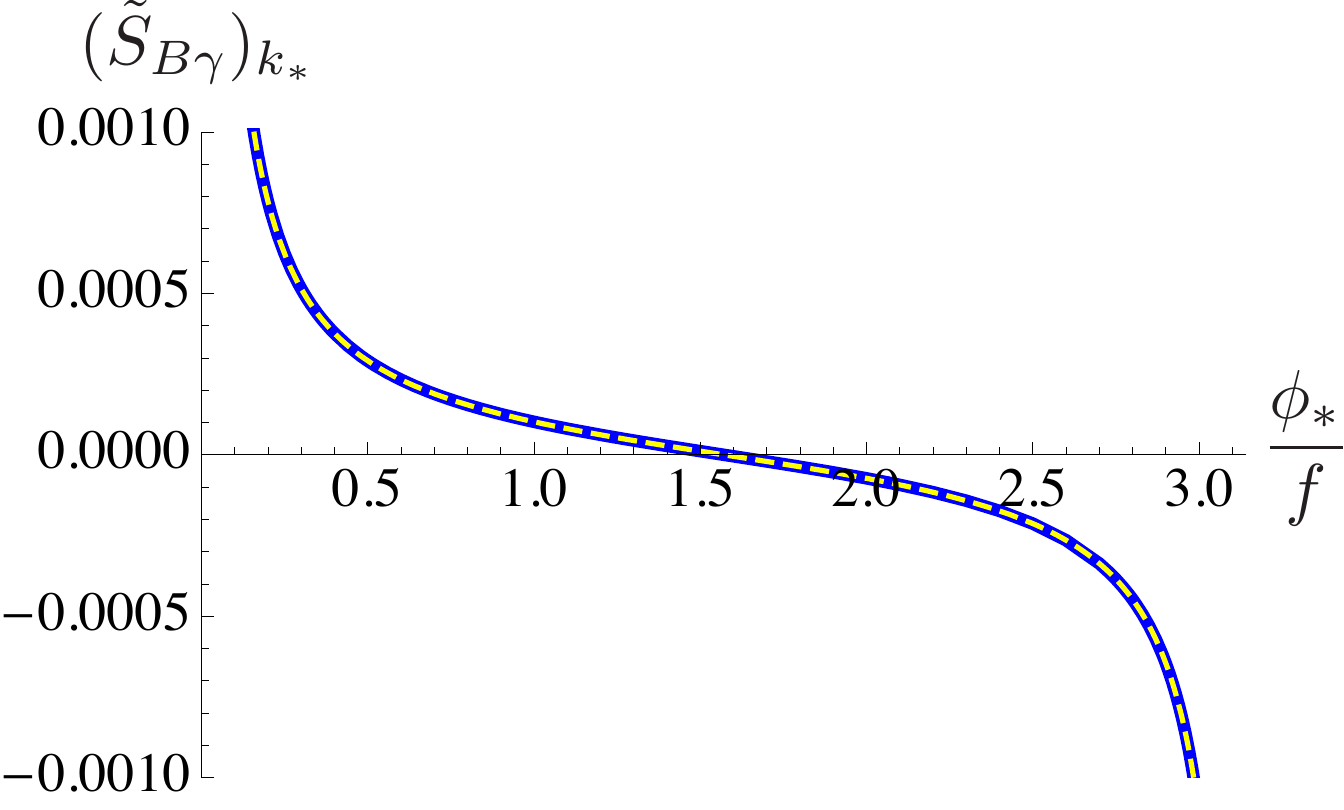}
  \end{center}
 \end{minipage} 
 \begin{minipage}{0.01\linewidth} 
  \begin{center}
  \end{center}
 \end{minipage} 
 \begin{minipage}{.48\linewidth}
  \begin{center}
 \includegraphics[width=\linewidth]{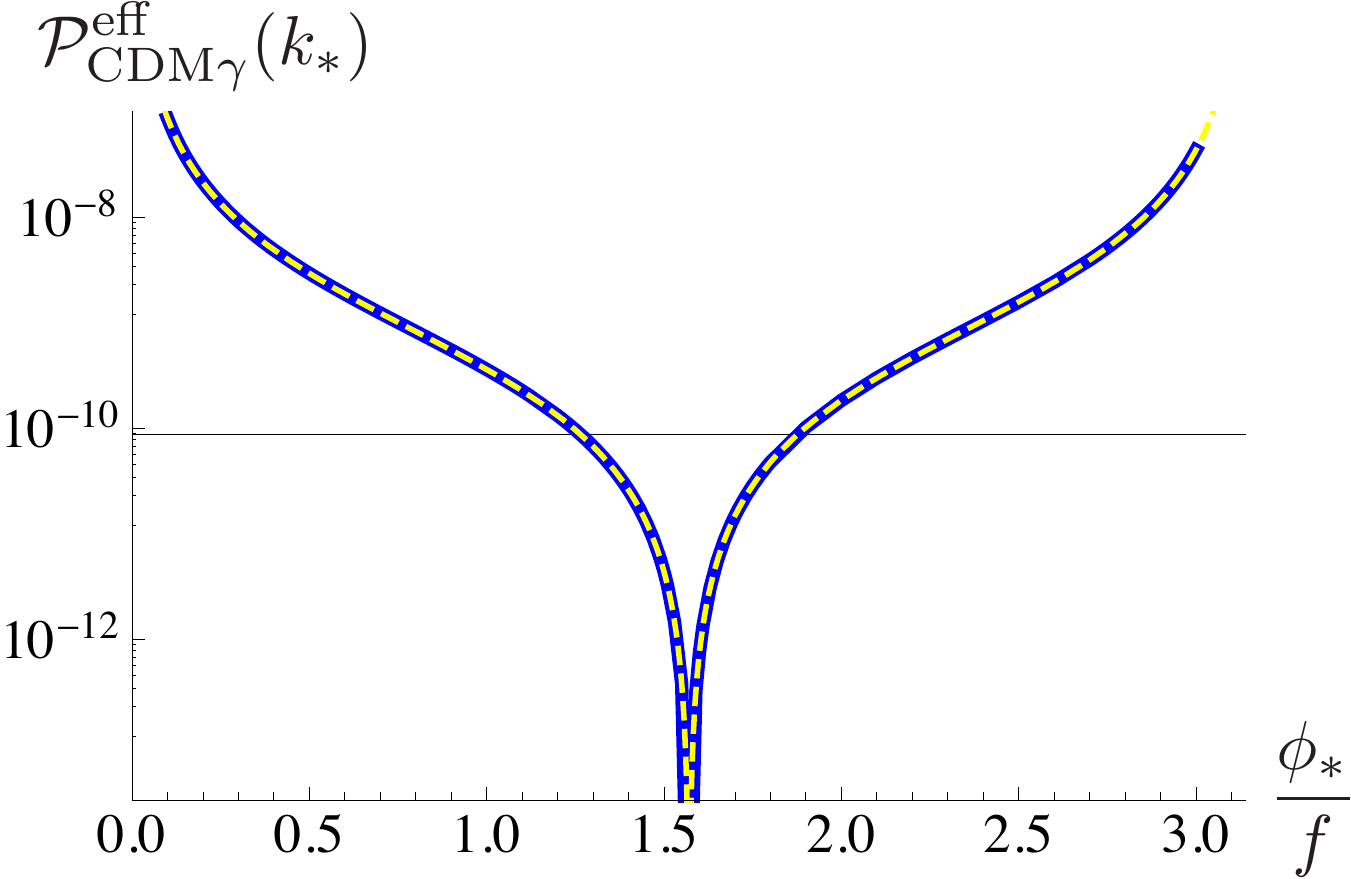}
  \end{center}
 \end{minipage} 
 \caption{Baryon isocurvature perturbation at the pivot scale, as a
 function of the initial field value along a cosine potential.
 Left: Fourier component amplitude of the baryon isocurvature.
 Right: Power spectrum amplitude of the effective CDM isocurvature.
 See the text for the precise definition of both quantities.
 Results from our analytic calculation Eq.~(\ref{eq7.5}) (yellow dashed)
 agree quite well with numerical computations (blue solid).
 The horizontal line in the right figure shows the {\it Planck} upper
 bound on the effective CDM isocurvature. The isocurvature perturbation
 is suppressed below the {\it Planck} bound for 
 field values within $1.3 \lesssim \phi_* / f \lesssim 1.9$.}
  \label{fig:SBI}
\end{figure}

For the computations, we used the parameters
\begin{equation}
   H_{*} = 1.0 \times 10^{12} \, \mathrm{GeV}, \thickspace
  H_{\mathrm{dec}} = 1.0 \times 10^{11}\, \mathrm{GeV}, \thickspace
 m_\phi = 2.2 \times 10^9\, \mathrm{GeV}, \thickspace
 f = 1.0 \times 10^{15} \, \mathrm{GeV}.
 \label{SBIpar}
\end{equation}
Note here that for a quadratic potential, 
this choice of $H_*$ and $f$ with $\abs{\phi_*} < \pi f$
would violate~(\ref{quad7}), producing too much baryon isocurvature. 
However this is not necessarily the case for the cosine potential.
In Figure~\ref{fig:SBI}, we show the resulting baryon isocurvature
perturbation at the pivot scale~$k_*$, 
as a function of the scalar position~$\phi_* / f$ when the mode~$k_*$
exits the horizon. The blue lines show the results from the numerical
computations, while the yellow dashed lines are from the analytic
result~(\ref{eq7.5}); the lines are on top of each other and thus one
sees that the two analyses agree quite well.

The left panel shows the Fourier mode amplitude~$ ( \tilde{S}_{B
\gamma})_{k_*}$.
Here we should note that, although the arbitrary phase which was fixed
in~(\ref{tildedeltaphih2pi}) does not affect physical observables,
the sign of the 
prefactor~$\tan^{-1}(\phi_{\mathrm{dec}}/f)$ in~$ ( \tilde{S}_{B
\gamma})_{k_*}$ does have a physical
meaning, as it specifies whether the $\phi$~velocity increases or
decreases for a larger field value.
This will be particularly important in the next subsection, when we
discuss compensated isocurvature perturbations.

In the right panel we display the effective CDM isocurvature
power~(\ref{effCDMiso}), and the {\it Planck} limit~(\ref{iso-con}) is
shown as the horizontal line. The isocurvature perturbation is seen to
fall below the {\it Planck} bound given that the initial field value
lies in the range of $1.3 \lesssim \phi_* / f \lesssim
1.9$.\footnote{Strictly speaking, the {\it Planck} limit of (\ref{iso-con}) is for a
scale-invariant isocurvature perturbation; if $\phi_*$ happens to lie
exactly where $ ( \tilde{S}_{B \gamma})_{k_*}$ crosses zero, then the
scale-dependence should be taken into account.}
For example, under the parameter set of~(\ref{SBIpar}),
together with one species~$i$ of $B_i = 1/3$, $c_i = -1$
(here a negative $c_i$ is chosen in order to create a positive Baryon
number at $ \phi_* > 0 $), $g_i = 2$,
with relativistic degrees of freedom $ g_* = g_{s*} = 106.75$,
and the decay rate parameter $\beta = 1$,
then one can check that the initial field value of 
$\phi_* / f \approx 1.5$ or $1.6$ satisfy all the conditions discussed
in Section~\ref{subsec:gencon}.
Note here that the $n_B/s$ normalization (\ref{nB-to-s}) gives two
solutions within the range $\abs{\phi_* }/f < \pi  $ for a cosine potential.
We should also mention that for the parameters given in~(\ref{SBIpar}),
the field excursion starting from $\phi_* / f \approx \pi /2$
during $10$~$e$-folds of inflation is as
tiny as $\abs{\Delta \phi} / f \sim 10^{-5}$; thus if the baryon isocurvature
is suppressed at the pivot scale, then it would also be suppressed over
the entire CMB scales.

Thus we have seen that the isocurvature constraint can be alleviated with
cosine potentials, allowing high-scale inflation to be compatible with
spontaneous baryogenesis. In particular, one can check that a parameter
window exists even 
with $H_*$ as high as $\sim 10^{14}\, \mathrm{GeV}$, which produces
large enough primordial gravitational waves to be observed in the near
future (or already been ruled out by current bounds on tensor perturbations.)
On the other hand, cosine potentials do not expand the window for
$m_\phi$ and $f$ as one would expect. This is because, as $\phi_*$ goes
beyond the inflection point, the
allowed window in the $m_\phi$-$f$ plane folds back and partially overlaps with
the window for $\phi_*$ in the quadratic region.

\subsection{Compensating Baryon Isocurvature with CDM Isocurvature}
\label{subsec:CIP}

If the scalar~$\phi$ is allowed to survive until the present, and
further if its abundance can be made comparable to that of CDM, then one
can expect $\phi$ to not only produce the baryons but also to serve as
CDM. 
As we have seen in the previous sections this possibility requires a
rather specific setup. 
In order to suppress the $\phi$~abundance,
for example, a large number of particles species~$i$
should be introduced.
As for $\phi$'s life time,
recall from the discussions below~(\ref{Gamma_phi}) that 
the $(\partial_\mu  \phi) j^\mu $ coupling typically provides a decay
channel into the $W$'s due to the chiral anomaly;
even if $\phi$ is lighter than $W$, it can still decay into
quarks through off-shell $W$'s. However it may be possible to suppress
the derivative coupling term with, e.g., a running coupling constant.

Despite the required tunings, a long-lived~$\phi$
has the benefit of, in addition to explaining dark matter,
being able to alleviate the isocurvature
constraint.
This is because the CDM consisting of~$\phi$ would also obtain 
isocurvature perturbations originating from the same field
fluctuations~$\delta \phi$ as the 
baryon isocurvature; thus it opens up the possibility of the baryon
isocurvature being compensated for by the CDM isocurvature. 
Such a perturbation is often referred to as a compensated
isocurvature perturbation, and is poorly constrained from observations
as it has no impact on the CMB at linear order.
For detailed discussions of current limits on compensated isocurvature,
see e.g.~\cite{Holder:2009gd,Grin:2011tf,Grin:2013uya},
which also discuss constraints from non-CMB probes as well as with
future experiments. 
In this section, we investigate the possibility of realizing compensated
isocurvature perturbation within spontaneous baryogenesis. 

\subsubsection{General Discussions of $\phi$ Density Isocurvature}

Let us for the moment assume that $\phi$ survives
until today and constitute (a fraction of) CDM, and compute its density
isocurvature perturbations:
\begin{equation}
 S_{\phi \gamma}  \equiv \frac{\delta \rho_\phi }{\rho_\phi }
  - \frac{3}{4} \frac{\delta \rho_\gamma }{\rho_\gamma }.
\end{equation}
Considering the $\phi$~density to redshift locally as $\rho_\phi \propto
a^{-3}$ since the onset of oscillations, we can proceed in a similar
fashion as we did for the baryon isocurvature in
Section~\ref{sec:baryon-iso}.
The main difference is that, instead of the decoupling surface, here we
need to consider the hypersurface of $H = H_{\mathrm{osc}}$ where
$\phi$ starts its oscillations.
However we should also remark that, unlike the decoupling surface, the
$H_{\mathrm{osc}}$-surface is generically not a uniform-density
slice; the simplest way to understand this is to note that the second
derivative of the scalar potential is not necessarily a constant.
The exceptional case is the quadratic potential~(\ref{quadV}),
where $H_{\mathrm{osc}}$ is set merely by the constant mass~$m_\phi$
(cf.~(\ref{Hosc-quad})) 
and thus the $H_{\mathrm{osc}}$-surface coincides with a
uniform-density slice.
However for nonquadratic potentials, the time when $\phi$ starts to oscillate
also depends on the field value itself, and therefore the field
experiences an inhomogeneous onset of oscillations~\cite{Kawasaki:2011pd}.
Hence for the $\phi$ density isocurvature perturbations, let us write 
\begin{equation}
 S_{\phi \gamma} =
  \left. \frac{\delta \rho_\phi }{\rho_\phi }  \right|_{H = H_{\mathrm{osc}}}
  + \cdots,
  \label{eq7.8}
\end{equation}
where ($\cdots$) represents the contribution to the $\phi$ density
fluctuation that arise when $H_{\mathrm{osc}}$ is space-dependent.
Here it should be noted that this extra contribution also originate from
the scalar field fluctuation obtained during inflation, hence should be
proportional to $\delta \phi$.  
On the other hand, 
the first term can be rewritten using $\rho_{\phi\, \mathrm{osc}} \simeq
V(\phi_{\mathrm{osc}})$ as
\begin{equation}
   \left. \frac{\delta \rho_\phi }{\rho_\phi }  \right|_{H =
    H_{\mathrm{osc}}} =
    \frac{V'(\phi_{\mathrm{osc}})}{V(\phi_{\mathrm{osc}})}
    \delta \phi_{\mathrm{osc}},
    \label{eq7.9}
\end{equation}
where $\delta \phi_{\mathrm{osc}}$ is the field fluctuation 
on the surface of $H= H_{\mathrm{osc}}$.
Therefore, moving to Fourier space and considering the
initial flat surface when the pivot scale~$k_*$ exits the horizon,
we can express the total $\phi$ density isocurvature perturbation as
\begin{equation}
 ({S}_{\phi  \gamma})_{ \bd{k}_* } =
  \left\{
    \frac{V'(\phi_{\mathrm{osc}})}{V(\phi_{\mathrm{osc}})} 
    \frac{\partial \phi_{\mathrm{osc}} }{\partial \phi_{*}}
+ \mathcal{X}
\right\}
  (\delta \phi_{*})_{ \bd{k}_*} .
  \label{S_phigamma}
\end{equation}
Here $\mathcal{X}$ is introduced to represent the perturbations
induced by the inhomogeneous onset of the oscillations.
$\mathcal{X}$ itself consists of unperturbed quantities, and can be
calculated analytically following the techniques developed for the
curvaton scenario in~\cite{Kawasaki:2011pd}.
However here we will not derive its explicit form and instead
numerically compute the isocurvature perturbations.
Nevertheless, the analytic form (\ref{S_phigamma}) will be useful for
obtaining a qualitative overview.
Using the field fluctuation amplitude~(\ref{phiH2pi}), we obtain a
general expression for the $\phi$~density isocurvature power spectrum,
\begin{equation}
 \mathcal{P}_{\phi \gamma} (k_*) =
\left\{
    \frac{V'(\phi_{\mathrm{osc}})}{V(\phi_{\mathrm{osc}})} 
    \frac{\partial \phi_{\mathrm{osc}} }{\partial \phi_{*}}
+ \mathcal{X}
\right\}^2
\left( \frac{H_*}{2 \pi} \right)^2.
  \label{P_phigamma}
\end{equation}

However, since the baryon and CDM ($\phi$) isocurvature
perturbations are hardly distinguished by CMB observations, 
we should refer to the two collectively as the effective CDM
isocurvature,
\begin{equation}
 S^{\mathrm{eff}}_{\mathrm{CDM} \gamma} = 
  \frac{\Omega_B}{\Omega_{\mathrm{CDM}}} 
  S_{B \gamma} 
    +  \frac{\Omega_\phi}{\Omega_{\mathrm{CDM}}} 
    S_{\phi \gamma},
        \label{effCDMiso2}
\end{equation}
whose power is bounded by {\it Planck} as~(\ref{iso-con}). 
Here, it is important to notice that the baryon isocurvature
perturbation~(\ref{SBgfourier}) and the $\phi$~density
perturbation~(\ref{S_phigamma}) both originate from the same field
fluctuation~$\delta \phi$, and thus they are perfectly
correlated.
Therefore the power spectrum of the effective CDM isocurvature is not a
simple sum of the individual power spectra (\ref{P_Bgamma}) and
(\ref{P_phigamma}), but instead is
\begin{equation}
 \mathcal{P}_{\mathrm{CDM} \gamma}^{\mathrm{eff}} (k_*) 
  = \left[
  \frac{\Omega_B}{\Omega_{\mathrm{CDM}}} 
  \frac{V''(\phi_{\mathrm{dec}})}{V'(\phi_{\mathrm{dec}})} 
  \frac{\partial \phi_{\mathrm{dec}} }{\partial \phi_{*}}
  + \frac{\Omega_\phi}{\Omega_{\mathrm{CDM}}} 
  \left\{
    \frac{V'(\phi_{\mathrm{osc}})}{V(\phi_{\mathrm{osc}})} 
    \frac{\partial \phi_{\mathrm{osc}} }{\partial \phi_{*}}
+ \mathcal{X}
       \right\}
  \right]^2
  \left( \frac{H_*}{2 \pi} \right)^2.
  \label{Peff7.13}
\end{equation}
Looking at this expression, one can imagine cases where the terms inside
the $[ \, \, ]$~parentheses cancel each other, thereby the baryon
isocurvature perturbation being compensated for by the CDM
isocurvature. 
When ignoring~$\mathcal{X}$ and assuming $\phi_*
\simeq \phi_{\mathrm{dec}} \simeq \phi_{\mathrm{osc}}$ for simplicity,
then it is further seen that the cancellation can happen only with
negatively curved potentials, i.e. $V'' < 0$. 
This can be understood as follows:
Since $V$ is by definition an increasing function of~$\abs{\phi -
\phi_{\mathrm{min}}}$, an initial position for~$\phi$ that is further
away from the potential minimum leads to a larger energy density at late
times, unless there is an overtaking.\footnote{This 
is also the reason why axion isocurvature perturbations can be enhanced
by anharmonic effects, but cannot be
eliminated~\cite{Lyth:1991ub,Kobayashi:2013nva}.} 
On the other hand, $\abs{\dot{\phi}}$ is set by the local tilt of the
potential, and thus increases with $\abs{\phi -
\phi_{\mathrm{min}}}$ when $V'' > 0$, and decreases when $V''< 0$. 
Thus with a negatively curved potential, the $\phi$~density isocurvature and baryon
isocurvature fluctuations have opposite signs.

\subsubsection{An Example with Compensated Isocurvature Perturbation}
\label{subsubsec:EX-CIP}

Now let us focus again on the cosine potential~(\ref{cosV}), and study
the resulting baryon and CDM ($\phi$) isocurvature perturbations, supposing
that $\phi$ survives until now.

Using the analytic expression~(\ref{Peff7.13}),
we can make a crude estimate of the isocurvature perturbations by
assuming $\phi_* \simeq \phi_{\mathrm{dec}} \simeq \phi_{\mathrm{osc}}$, and
further ignoring effects from the inhomogeneous onset of
oscillations~$\mathcal{X}$.
Then, in terms of the Fourier mode amplitude defined 
as~(\ref{Ssplit}), one obtains
\begin{equation}
 ( \tilde{S}^{\mathrm{eff}}_{\mathrm{CDM} \gamma})_{k_*} \sim 
  \left\{
   \frac{\Omega_{B}}{\Omega_{\mathrm{CDM}}} 
   \tan^{-1} \left(\frac{\phi_{*}}{f} \right)
   +    \frac{\Omega_\phi}{\Omega_{\mathrm{CDM}}} 
    \tan^{-1} \left(\frac{\phi_{*}}{2 f} \right)
	  \right\}
  \frac{H_*}{2 \pi f}.
  \label{Seff-crude}
\end{equation}
We can expect this approximation to work well at small $\abs{\phi_*} /
f$, where the potential approaches a quadratic and thus $\mathcal{X}$ is
actually absent.

We have also numerically computed the isocurvature perturbations, as was
done in Section~\ref{subsec:SBI}. However here we also computed the
$\phi$~density isocurvature perturbations via
$S_{\phi \gamma} = (\delta \rho_{\phi} / \rho_\phi)_{\rho =
\mathrm{const.}}$,
by evaluating the density fluctuations
on some arbitrary uniform-density slice when $\phi$ is harmonically
oscillating. 

The model parameters here should be chosen to suppress the
$\phi$~abundance as $\Omega_\phi \leq \Omega_{\mathrm{CDM}}$,
so that $\phi$ can be the CDM.
The analysis of $\Omega_\phi$ for the quadratic
case~(\ref{quadOmegaphi}) indicates that, for instance, a number of
particle species~$i$ greater than $\sim 100$
can be used for suppressing~$\Omega_\phi$.
Recall also that the bound of~(\ref{quadOmegaphi}) was derived
using the lower bound~(\ref{quad2-4}) on~$H_{\mathrm{dec}}$,
thus a low decoupling scale is further needed.
(Hence, while the scenario with compensated isocurvature 
can alleviate the isocurvature constraint,
it will typically still disfavor high-scale inflation.)
For the numerical computations, we chose 
\begin{equation}
   H_{*} = 1.0 \times 10^{7} \, \mathrm{GeV}, \thickspace
  H_{\mathrm{dec}} = 1.0 \times 10^{-2}\, \mathrm{GeV}, \thickspace
 m_\phi = 1.3 \times 10^{-2}\, \mathrm{GeV}, \thickspace
 f = 3.2 \times 10^{9} \, \mathrm{GeV}.
 \label{CIPpar}
\end{equation}
Such a choice, especially with the large hierarchy between
$H_{*}$ and $H_{\mathrm{dec}}$, was prohibited 
in Section~\ref{subsec:CS-quad} where we considered a `reasonable'
set of particle species. However here we are interested in the extreme
case with $\Omega_\phi \leq \Omega_{\mathrm{CDM}}$,
which is allowed only under specific setups in the particle content,
such as with a large number of species. 
By fine tuning the parameters in the particle content,
it is in principle possible to adopt the parameters of (\ref{CIPpar}).
We also note that with this parameter set,
if $\phi$ is a PNGB with $f$ being the symmetry breaking scale,
then the reheating temperature needs to be close to the decoupling
temperature to avoid the symmetry from being restored at reheating.

\begin{figure}[t!]
 \begin{minipage}{.48\linewidth}
  \begin{center}
 \includegraphics[width=\linewidth]{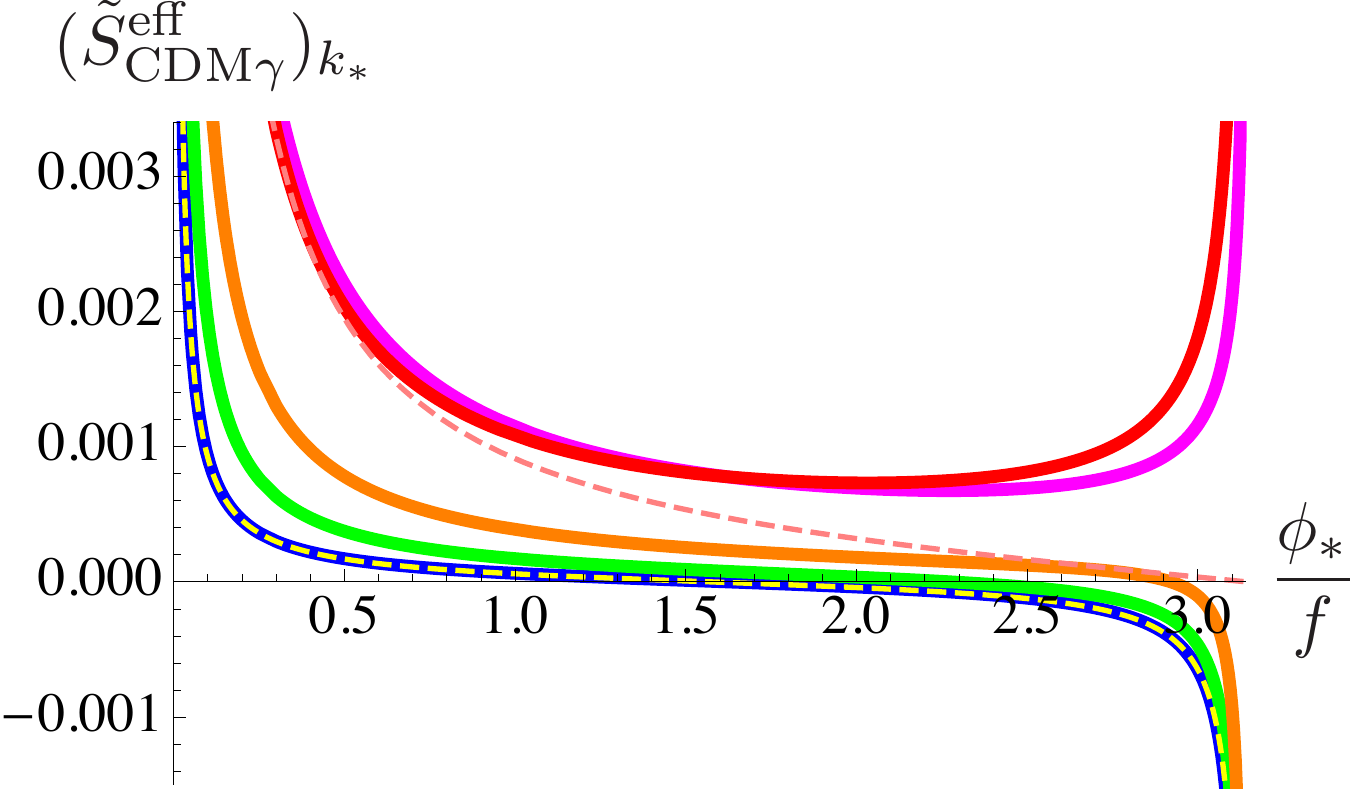}
  \end{center}
 \end{minipage} 
 \begin{minipage}{0.01\linewidth} 
  \begin{center}
  \end{center}
 \end{minipage} 
 \begin{minipage}{.48\linewidth}
  \begin{center}
 \includegraphics[width=\linewidth]{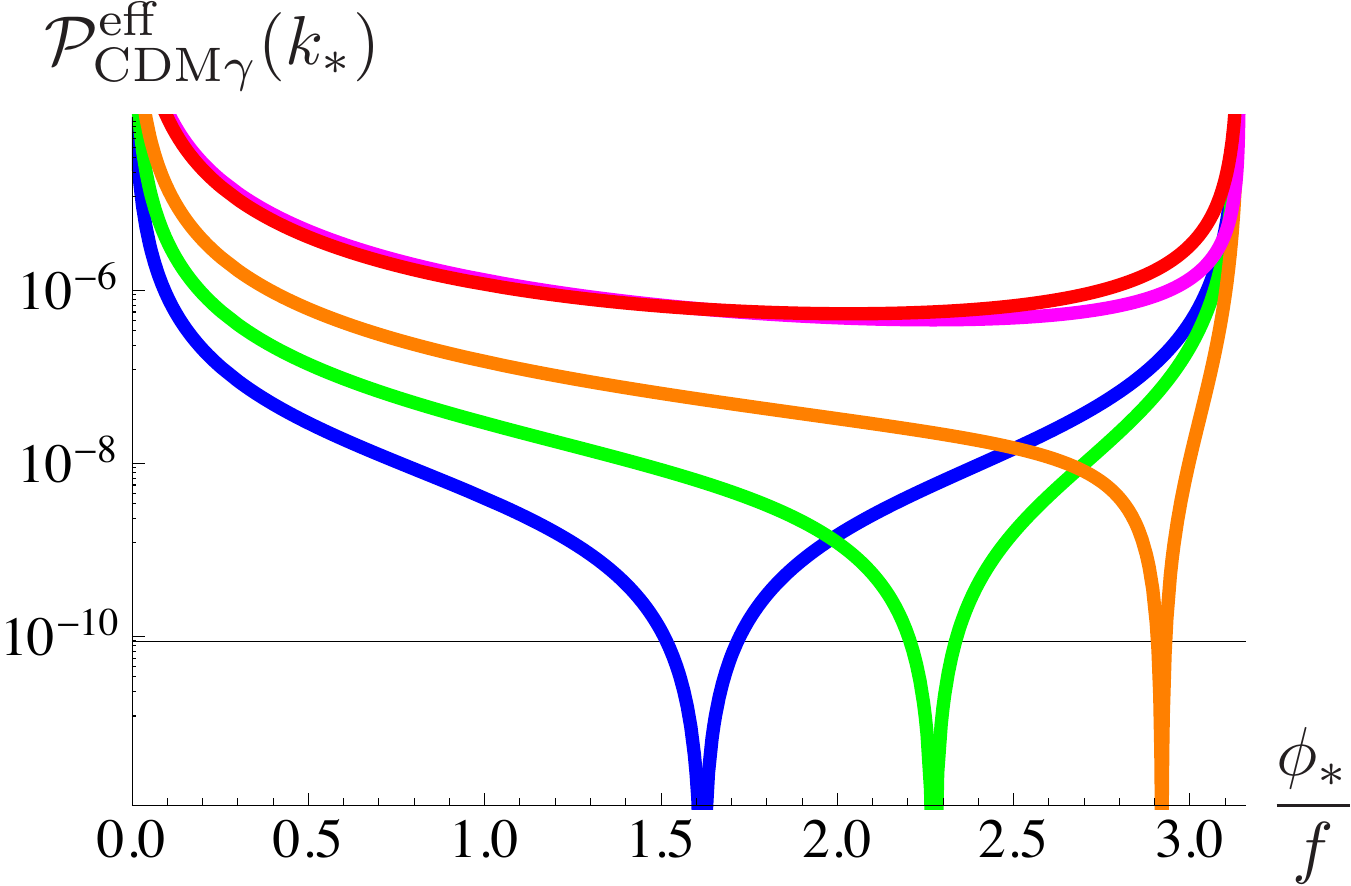}
  \end{center}
 \end{minipage} 
 \caption{Baryon and CDM isocurvature perturbations at the pivot scale, as
 functions of the initial field value along a cosine potential.
 Left: Fourier component amplitudes.
 Right: Power spectrum amplitudes.
 See the text for the precise definition of both quantities.
 The solid lines show numerically computed results for
 the normalized baryon isocurvature~$ (\Omega_B/\Omega_\mathrm{CDM})
 S_{B \gamma}$ (blue),
 $\phi$~density isocurvature~$ S_{\phi \gamma}$ (red), and 
 effective CDM isocurvature~$ S_{\mathrm{CDM} \gamma}^{\mathrm{eff}}$ with
 $\Omega_\phi/\Omega_\mathrm{CDM} = 0.1$ (green), $0.3$ (orange), $1$
 (magenta).
 In the left panel we also show the analytic estimates of the normalized
 baryon isocurvature (yellow dashed), and $\phi$~density isocurvature
 (pink dashed).
 The horizontal line in the right figure shows the {\it Planck} upper
 bound on the effective CDM isocurvature.
 The baryon isocurvature is compensated for by CDM isocurvature
 in the region $\phi_* / f > \pi /2$,
 where the {\it Planck} constraint can be evaded even though the individual
 baryon and CDM isocurvature perturbations are large.}
  \label{fig:CIP}
\end{figure}

We show the resulting baryon and CDM ($\phi$) isocurvature perturbations
at the pivot scale in Figure~\ref{fig:CIP}, where the amplitudes of the
Fourier mode~$(\tilde{S})_{k_*}$ and the power
spectrum~$\mathcal{P}(k_*)$ are shown in the left and right panels,
respectively.
The numerically computed results are shown as the solid lines. 
The blue lines represent the baryon isocurvature~$S_{B \gamma}$
normalized by the abundance ratio~$\Omega_B / \Omega_{\mathrm{CDM}}$,
and the red lines are for the $\phi$~density isocurvature~$S_{\phi
\gamma}$ without any normalization factor.
In the left panel we also show the crude estimates given by the analytic
expression of~(\ref{Seff-crude}), whose first term gives the normalized
baryon isocurvature 
(shown as the yellow dashed line in the figure), and the second term
without $\Omega_\phi / \Omega_{\mathrm{CDM}}$ gives the
$\phi$~density isocurvature (pink dashed line).
As for the baryon isocurvature, the numerical (blue) and analytic (yellow
dashed) results are seen to agree quite well, as was also the case in
Figure~\ref{fig:SBI}.
On the other hand for the $\phi$~density isocurvature, as we have
expected, the numerical (red) and analytic (pink dashed) results agree
well only at $\phi_* / f < 1$.
While the analytic estimate asymptotes to zero as $\phi_* $ approaches
the hilltop (i.e.~$\pi f$), the numerical results turn into an
increasing function above the inflection point.
In particular, the $\phi$~density fluctuations are strongly enhanced
close to the hilltop.
These behaviors arise due to the inhomogeneous onset of oscillations,
which is known to produce large perturbations especially for hilltop potentials.
Effects of the inhomogeneous onset of oscillations have
been studied both numerically 
and analytically in previous works in the context of
curvatons~\cite{Kawasaki:2011pd,Kawasaki:2012gg}, and QCD 
axions~\cite{Lyth:1991ub,Kobayashi:2013nva}.

In the figures we also display the effective CDM
isocurvature~$S^{\mathrm{eff}}_{\mathrm{CDM} \gamma}$~(\ref{effCDMiso2}).
In order to see how the presence of the $\phi$~density isocurvature
affects the total effective isocurvature,
here we have taken the $\phi$~density fraction~$\Omega_\phi /
\Omega_{\mathrm{CDM}}$ as a free parameter;\footnote{Here we are treating
the fraction~$\Omega_\phi / \Omega_{\mathrm{CDM}}$ as a free parameter,
however we should note that the $\phi$~abundance actually depends
on~$\phi_*$. The abundance increases especially as $\phi_*$ moves
towards the hilltop, as the onset of the oscillation is delayed.}
we have chosen $\Omega_\phi / \Omega_{\mathrm{CDM}} = 0.1$, $0.3$, and $1$, 
which are respectively shown as the green, orange, and magenta lines.
As $\Omega_\phi / \Omega_{\mathrm{CDM}}$ is varied between $0$ and $1$, 
the effective CDM isocurvature shifts between the normalized baryon
isocurvature and the $\phi$~density isocurvature.
The effective isocurvature with $\Omega_\phi / \Omega_{\mathrm{CDM}} =
0$, i.e. the pure baryon isocurvature (blue), is suppressed at around the
inflection point~$\phi_* / f = \pi /2$,
as we already studied in Section~\ref{subsec:SBI}.
For nonzero values of~$\Omega_\phi / \Omega_{\mathrm{CDM}}$,
the baryon isocurvature is compensated for by CDM isocurvature
in the region $\phi_* / f > \pi /2$.
As $\Omega_\phi / \Omega_{\mathrm{CDM}}$ increases,
the value of~$\phi_*$ where $S^{\mathrm{eff}}_{\mathrm{CDM} \gamma}$ crosses
zero shifts towards larger values; there the isocurvature constraints
from CMB measurements are evaded, even though the individual baryon and
CDM isocurvature perturbations are large. 
At $\Omega_\phi / \Omega_{\mathrm{CDM}} \gtrsim 0.8 $,
the effective CDM isocurvature closely follows the $\phi$~density
isocurvature, hence is nonzero for all values of~$\phi_*$.

\section{Conclusions}
\label{sec:conc}

The goal of this paper was to investigate the cosmological aspects of
spontaneous baryogenesis driven by a scalar field.
We have provided general requirements for a successful spontaneous
baryogenesis that are independent of the particle physics model.
The particularly important constraints were obtained by studying the
backreaction of the produced baryons
on the scalar field during baryogenesis, 
the cosmological expansion history after baryogenesis in the presence of the
oscillating scalar, and the baryon isocurvature perturbations.  
The various constraints are summarized in Section~\ref{subsec:gencon}.

We then performed a comprehensive study of the minimal scenario with a
quadratic scalar potential, and demonstrated that cosmological
considerations alone tightly restrict the model parameters.
It was shown that the energy scales of inflation, reheating,
and decoupling cannot be separated by more than a few orders of
magnitude, and thus an efficient (p)reheating is required for
spontaneous baryogenesis. Furthermore, for a given inflation scale, the
mass and the decay constant of the scalar are constrained to lie within
ranges of at most a few orders of magnitude. 
This in turn suggests that, once any of the model parameters is fixed by
other considerations such as from particle physics, then the inflation
scale can be predicted. 
As the minimal scenario is thus tightly constrained from cosmology,
it would be interesting to explore its phenomenological consequences.
It is also very important to study explicit
constructions of spontaneous baryogenesis in particle physics models,
taking into account our cosmological constraints. 
We also note that the generalization of our discussions to 
scenarios of baryogenesis via spontaneous leptogenesis is straightforward.

As an extension to the minimal setup, we further explored
spontaneous baryogenesis with a nonquadratic scalar potential.
We particularly focused on periodic potentials with inflection points,
such as a cosine potential,
which can arise if the scalar is a PNGB of an approximate U(1) symmetry
corresponding to the baryon number.
We showed that the baryon isocurvature perturbation vanishes in the
vicinity of the inflection point, and thus the tension between
high-scale inflation and spontaneous baryogenesis can be alleviated.

We also explored a possibility that the scalar survives until now and
constitutes (a fraction of) CDM. We explicitly demonstrated that in such cases,
the baryon isocurvature perturbation can be compensated by the CDM
isocurvature, and therefore escapes the CMB constraints. 

One of the general lessons of this work is that any scenario
that exploits scalar condensates in the early universe
can leave non-negligible traces in the subsequent cosmology, therefore
requires careful considerations. 
Cosmological constraints on such scenarios are especially powerful when
the scalar field dynamics in the very early times is related to that in
later times in a rather straightforward way.  
On the other hand, the presence of a strongly
time-dependent scalar potential, or a strong renormalization group
running, or multi-field dynamics can complicate the relation between the
physics at early and late times.
It would also be interesting to extend our analysis to study such
cosmological scenarios.

\section*{Acknowledgments}

It is a pleasure to thank Paolo Creminelli, Daniel Grin, Stefano
Liberati, Atsushi Naruko, Serguey Petcov, Alexander J. Stuart, Fuminobu
Takahashi, and Alexander van Engelen for helpful conversations.
T.K. acknowledges support from the INFN INDARK initiative.


\appendix

\section{Comments on Spontaneous Baryogenesis from Decay of $\phi$}
\label{app:A}

In the main text, we have analyzed spontaneous baryogenesis driven by
a scalar field that slowly varies along its potential
while baryon violating interactions are in
thermal equilibrium. However, as was pointed out in~\cite{Cohen:1987vi},
there can also be baryogenesis after the interactions have decoupled,
when the scalar decays through the $(\partial_\mu \phi) j^\mu$ coupling
into particles that carry baryon numbers. 
The works~\cite{Dolgov:1994zq,Dolgov:1996qq} further studied this effect
in a flat spacetime, by treating the decaying~$\phi$ as a classical
field with damped oscillations.
Through Bogoliubov calculations, they found the net baryon
density produced by the time the oscillation has damped away to be
proportional to the cube of the initial field amplitude, 
\begin{equation}
 \abs{n_B} \sim \Gamma_\phi f^2 \left( \frac{\bar{\phi}}{f} \right)^3,
  \label{A.1}
\end{equation}
where $\bar{\phi}$ is the initial oscillation amplitude of~$\phi$.
Here, whether a net baryon or antibaryon number is produced is set by
the initial phase of the oscillation. 

We can expect the result (\ref{A.1}) in flat space to be generalized
to an expanding universe by 
taking $\bar{\phi}$ as the oscillation amplitude right before the
decay of~$\phi$, i.e. when $H = \Gamma_\phi$.
Here, as $\phi$ already begins to slowly decay once it starts to
oscillate, one may expect to use the field amplitude at the onset of
oscillation $\abs{\phi_{\mathrm{osc}}}$, instead of at later when
$H = \Gamma_\phi$.
However it should be noted that while $H \gg \Gamma_\phi$, the change in the
oscillation amplitude due to the decay of~$\phi$ during one Hubble time
is suppressed as
\begin{equation}
 | \Delta \bar{\phi}^3 | \sim
   \bar{\phi}^2 | \Delta \bar{\phi} | \sim
   \frac{\Gamma_\phi}{H}\bar{\phi}^3 .
   \label{A.2}
\end{equation}
Thus during the early stages of the oscillation, the field
amplitude is damped mainly due to the expansion of the universe.
In other words, in an expanding universe,
only a tiny fraction of the $\phi$~density at the onset
of the oscillations can be used for creating 
baryons (unless $\phi$ decays soon
after starting to oscillate.)
Therefore one cannot simply substitute $\abs{\phi_{\mathrm{osc}}}$
into~(\ref{A.1}) to estimate the baryon number produced
from the oscillating scalar while $H \gg \Gamma_\phi$.
The baryon production at the beginning of the
oscillations is expected to be suppressed due to the factor of
$\Gamma_\phi / H$, but it would be worthwhile to compute this effect
explicitly by including the expansion of the universe in the
calculations.  

Let us now give an order-of-magnitude estimate of the baryon asymmetry
produced from the scalar decay,
by using (\ref{A.1}) with $\bar{\phi}$ taken to be the
oscillation amplitude when $ H = \Gamma_\phi$, i.e.,
\begin{equation}
 \abs{n_B} \sim \Gamma_\phi f^2
  \left( \frac{\bar{\phi}_{\mathrm{decay}}}{f} \right)^3.
  \label{A.3}
\end{equation}
Here we focus on the case with a quadratic potential for the scalar, 
\begin{equation}
 V(\phi) = \frac{1}{2} m_\phi^2 \phi^2,
\end{equation}
under which the field amplitude of an oscillating~$\phi$ redshifts as
$\bar{\phi} \propto a^{-3/2}$. 
In this appendix we are interested in spontaneous baryogenesis induced by the decay
of~$\phi$, thus we do not necessarily have to require reheating to
have taken place prior to the decay
(although, the oscillation still should start after
inflation, otherwise the field amplitude would be damped away and the
produced baryon number~(\ref{A.3}) would be extremely tiny.)
However we will soon see that in a
matter-dominated universe, the resulting baryon asymmetry would be
smaller compared to that in a radiation-dominated universe.
Hence let us consider a radiation-dominated
background and use $H \propto a^{-2}$,
where for simplicity we ignore the time-variation of the relativistic
degrees of freedom.
Then the field amplitude upon decay is expressed in terms of the
field value at the onset of oscillations as
\begin{equation}
 \bar{\phi}_{\mathrm{decay}} = \abs{\phi_{\mathrm{osc}}}
  \left( \frac{ \Gamma_\phi  }{H_{\mathrm{osc}}} \right)^{3/4} .
  \label{A.5}
\end{equation}
Further using $H_{\mathrm{osc}} \sim m_\phi$,
$\phi_{\mathrm{osc}} \sim \phi_*$ (as in the main text,
asterisks are used for values when the pivot scale~$k_*$ exits the
horizon during inflation), and also $s_{\mathrm{decay}} \sim (M_p \Gamma_\phi)^{3/2}$ for the
entropy density at $\phi$-decay, we obtain the baryon-to-photon ratio as
\begin{equation}
 \left| \frac{n_B}{s} \right|_{\mathrm{decay}} \sim 
  \frac{f^2}{M_p^{3/2} \Gamma_{\phi}^{1/2}}
  \left( \frac{\abs{\phi_*}}{f} \right)^3
  \left( \frac{\Gamma_\phi }{m_\phi } \right)^{9/4}
  \sim \frac{\beta^{7/4} \, m_\phi^3}{ M_p^{3/2} \abs{\phi_*}^{3/2}}
\left( \frac{\abs{\phi_*}}{f}\right)^{9/2} .
  \label{app-a}
\end{equation}
Upon moving to the far right hand side, we have parameterized the
decay rate in terms of~$\beta$ as we did in~(\ref{Gamma_phi}). 

The baryon isocurvature perturbation is estimated as
$\mathcal{P}_{B \gamma} (k_*) \sim (H_* / 2 \pi \phi_*)^2$,
hence the {\it Planck} limit on isocurvature~(\ref{iso-con}) yields
$ ( H_*/\phi_* )^2 \lesssim 10^{-7}$.
Further using $m_\phi < H_*$,
the field bound $\abs{\phi_*} \lesssim f$ (cf. discussions
around~(\ref{field-range})),
and $ \beta \lesssim 1$ (cf. below~(\ref{Gamma_phi})), 
we obtain an upper bound on the ratio~(\ref{app-a}) as
\begin{equation}
 \left| \frac{n_B}{s} \right|_{\mathrm{decay}} \lesssim
  10^{-5} \left( \frac{H_*}{M_p} \right)^{3/2}.
\end{equation}
As current observational limits on primordial gravitational waves
indicate an upper bound on the inflation scale of
$H_* \lesssim 10^{14}\, \mathrm{GeV}$~\cite{Ade:2015lrj}, 
the baryon asymmetry produced at the decay is bounded as
\begin{equation}
 \left| \frac{n_B}{s} \right|_{\mathrm{decay}} \lesssim 10^{-12}.
\end{equation}
This bound becomes stricter if the $\phi$-decay happens during an 
(effectively) matter-dominated epoch (e.g. while the universe is
dominated by an oscillating inflaton, or if $\phi$ comes to dominate the
universe before decaying), 
as then the power of the suppression factor $\Gamma_\phi /
H_{\mathrm{osc}}$ in~(\ref{A.5}) would become larger.
We also mention that the produced baryon asymmetry is 
smaller if $\phi$ also has decay channels into particles without a
baryon number. 
Thus we conclude that, at least with quadratic potentials,
spontaneous baryogenesis induced by the decay of~$\phi$ is
insignificant.



\begin{thebibliography}{99}

\bibitem{Sakharov:1967dj} 
  A.~D.~Sakharov,
  Pisma Zh.\ Eksp.\ Teor.\ Fiz.\  {\bf 5}, 32 (1967)
  [JETP Lett.\  {\bf 5}, 24 (1967)]
  [Sov.\ Phys.\ Usp.\  {\bf 34}, 392 (1991)]
  [Usp.\ Fiz.\ Nauk {\bf 161}, 61 (1991)].
 
\bibitem{Cohen:1987vi} 
  A.~G.~Cohen and D.~B.~Kaplan,
  Phys.\ Lett.\ B {\bf 199}, 251 (1987).

\bibitem{Cohen:1988kt} 
  A.~G.~Cohen and D.~B.~Kaplan,
  Nucl.\ Phys.\ B {\bf 308}, 913 (1988).

\bibitem{Dine:1990fj} 
  M.~Dine, P.~Huet, R.~L.~Singleton, Jr and L.~Susskind,
  Phys.\ Lett.\ B {\bf 257}, 351 (1991).

\bibitem{Cohen:1991iu} 
  A.~G.~Cohen, D.~B.~Kaplan and A.~E.~Nelson,
  Phys.\ Lett.\ B {\bf 263}, 86 (1991).

\bibitem{Dolgov:1994zq} 
  A.~Dolgov and K.~Freese,
  Phys.\ Rev.\ D {\bf 51}, 2693 (1995)
  [hep-ph/9410346].

\bibitem{Dolgov:1996qq} 
  A.~Dolgov, K.~Freese, R.~Rangarajan and M.~Srednicki,
  Phys.\ Rev.\ D {\bf 56}, 6155 (1997)
  [hep-ph/9610405].

\bibitem{Li:2001st} 
  M.~z.~Li, X.~l.~Wang, B.~Feng and X.~m.~Zhang,
  Phys.\ Rev.\ D {\bf 65}, 103511 (2002)
  [hep-ph/0112069].
	
\bibitem{Chiba:2003vp} 
  T.~Chiba, F.~Takahashi and M.~Yamaguchi,
  Phys.\ Rev.\ Lett.\  {\bf 92}, 011301 (2004)
  Erratum: [Phys.\ Rev.\ Lett.\  {\bf 114}, no. 20, 209901 (2015)]
  [hep-ph/0304102].

\bibitem{Carroll:2005dj} 
  S.~M.~Carroll and J.~Shu,
  Phys.\ Rev.\ D {\bf 73}, 103515 (2006)
  [hep-ph/0510081].
	
\bibitem{Kusenko:2014lra} 
  A.~Kusenko, L.~Pearce and L.~Yang,
  Phys.\ Rev.\ Lett.\  {\bf 114}, no. 6, 061302 (2015)
  [arXiv:1410.0722 [hep-ph]].

\bibitem{Kusenko:2014uta} 
  A.~Kusenko, K.~Schmitz and T.~T.~Yanagida,
  Phys.\ Rev.\ Lett.\  {\bf 115}, no. 1, 011302 (2015)
  [arXiv:1412.2043 [hep-ph]].

\bibitem{Daido:2015gqa} 
  R.~Daido, N.~Kitajima and F.~Takahashi,
  JCAP {\bf 1507}, no. 07, 046 (2015)
  [arXiv:1504.07917 [hep-ph]].

\bibitem{Starobinsky:1980te} 
  A.~A.~Starobinsky,
  Phys.\ Lett.\ B {\bf 91}, 99 (1980).

\bibitem{Sato:1980yn} 
  K.~Sato,
  Mon.\ Not.\ Roy.\ Astron.\ Soc.\  {\bf 195}, 467 (1981).

\bibitem{Guth:1980zm} 
  A.~H.~Guth,
  Phys.\ Rev.\ D {\bf 23}, 347 (1981).
	
\bibitem{Turner:1988sq} 
  M.~S.~Turner, A.~G.~Cohen and D.~B.~Kaplan,
  Phys.\ Lett.\ B {\bf 216}, 20 (1989).

\bibitem{Davoudiasl:2004gf} 
  H.~Davoudiasl, R.~Kitano, G.~D.~Kribs, H.~Murayama and P.~J.~Steinhardt,
  Phys.\ Rev.\ Lett.\  {\bf 93}, 201301 (2004)
  [hep-ph/0403019].

\bibitem{Ade:2015xua} 
  P.~A.~R.~Ade {\it et al.} [Planck Collaboration],
  arXiv:1502.01589 [astro-ph.CO].

\bibitem{Chiba:2009sj} 
  T.~Chiba,
  Phys.\ Rev.\ D {\bf 79}, 083517 (2009)
  Erratum: [Phys.\ Rev.\ D {\bf 80}, 109902 (2009)]
  [arXiv:0902.4037 [astro-ph.CO]].

\bibitem{Kawasaki:2011pd} 
  M.~Kawasaki, T.~Kobayashi and F.~Takahashi,
  Phys.\ Rev.\ D {\bf 84}, 123506 (2011)
  [arXiv:1107.6011 [astro-ph.CO]].

\bibitem{Linde:1996gt} 
  A.~D.~Linde and V.~F.~Mukhanov,
  Phys.\ Rev.\ D {\bf 56}, 535 (1997)
  [astro-ph/9610219].

\bibitem{Enqvist:2001zp} 
  K.~Enqvist and M.~S.~Sloth,
  Nucl.\ Phys.\ B {\bf 626}, 395 (2002)
  [hep-ph/0109214].

\bibitem{Lyth:2001nq} 
  D.~H.~Lyth and D.~Wands,
  Phys.\ Lett.\ B {\bf 524}, 5 (2002)
  [hep-ph/0110002].

\bibitem{Moroi:2001ct} 
  T.~Moroi and T.~Takahashi,
  Phys.\ Lett.\ B {\bf 522}, 215 (2001)
  [Phys.\ Lett.\ B {\bf 539}, 303 (2002)]
  [hep-ph/0110096].

\bibitem{Lyth:2009zz} 
  D.~H.~Lyth and A.~R.~Liddle,
  {\it The primordial density perturbation: Cosmology, inflation and the origin of structure,}
  Cambridge University Press, 2009.
	
\bibitem{Ade:2015lrj} 
  P.~A.~R.~Ade {\it et al.} [Planck Collaboration],
  arXiv:1502.02114 [astro-ph.CO].

\bibitem{Silverstein:2008sg} 
  E.~Silverstein and A.~Westphal,
  Phys.\ Rev.\ D {\bf 78}, 106003 (2008)
  [arXiv:0803.3085 [hep-th]].

\bibitem{McAllister:2008hb} 
  L.~McAllister, E.~Silverstein and A.~Westphal,
  Phys.\ Rev.\ D {\bf 82}, 046003 (2010)
  [arXiv:0808.0706 [hep-th]].
	
\bibitem{Holder:2009gd} 
  G.~P.~Holder, K.~M.~Nollett and A.~van Engelen,
  Astrophys.\ J.\  {\bf 716}, 907 (2010)
  [arXiv:0907.3919 [astro-ph.CO]].

\bibitem{Grin:2011tf} 
  D.~Grin, O.~Dore and M.~Kamionkowski,
  Phys.\ Rev.\ D {\bf 84}, 123003 (2011)
  [arXiv:1107.5047 [astro-ph.CO]].

\bibitem{Grin:2013uya} 
  D.~Grin, D.~Hanson, G.~P.~Holder, O.~Doré and M.~Kamionkowski,
  Phys.\ Rev.\ D {\bf 89}, no. 2, 023006 (2014)
  [arXiv:1306.4319 [astro-ph.CO]].

\bibitem{Lyth:1991ub} 
  D.~H.~Lyth,
  Phys.\ Rev.\ D {\bf 45}, 3394 (1992).

\bibitem{Kobayashi:2013nva} 
  T.~Kobayashi, R.~Kurematsu and F.~Takahashi,
  JCAP {\bf 1309}, 032 (2013)
  [arXiv:1304.0922 [hep-ph]].
	
\bibitem{Kawasaki:2012gg} 
  M.~Kawasaki, T.~Kobayashi and F.~Takahashi,
  JCAP {\bf 1303}, 016 (2013)
  [arXiv:1210.6595 [astro-ph.CO]].
\end{thebibliography}
\end{document}